\documentclass[twocolumn]{aastex631}

\received{\today}

\submitjournal{ApJ}

\usepackage {threeparttable}
\usepackage{amsmath}

\shorttitle{Off-axis GRB afterglow in SN radio-lightcurve}
\shortauthors{Kusafuka, Matsuoka \& Sawada}

\graphicspath{{./}{figures/}}
\graphicspath{{./}{old/}}

\begin{document}


\title{Unifying Circumstellar Environment in Broad-Lined Type Ic Radio Supernovae Towards Off-axis Gamma-Ray Burst Exploration}


\author[0000-0001-5963-4436]{Yo Kusafuka}
\affiliation{Institute for Cosmic Ray Research, The University of Tokyo, Kashiwa, Chiba 277-8582, Japan}
\correspondingauthor{Yo Kusafuka}
\email{kusafuka@icrr.u-tokyo.ac.jp}

\author[0000-0002-6916-3559]{Tomoki Matsuoka}
\affiliation{Institute of Astronomy and Astrophysics, Academia Sinica, No.1, Sec. 4, Roosevelt Road, Taipei 106319, Taiwan}

\author[0000-0003-4876-5996]{Ryo Sawada}
\affiliation{Institute for Cosmic Ray Research, The University of Tokyo, Kashiwa, Chiba 277-8582, Japan}

\begin{abstract}

Decades have passed since the first confirmed association between a broad-lined Type Ic supernova (Type IcBL SN) and a long gamma-ray burst (GRB), and the number of known GRB–SN associations has steadily increased. However, it is important to note that the radiation from GRB afterglows and the radio emission from SNe may be both produced by outflows evolving within the same ambient medium.
In this study, we present the first comprehensive theoretical predictions of radio emission from a Type IcBL supernova associated with a GRB jet, explicitly accounting for the structure of the ambient medium. We model each component of the radio emission, with particular emphasis on exploring wide ranges of isotropic explosion energy and viewing angle in our GRB afterglow calculations. We show that, within specific regions of parameter space, the composite radio light curve exhibits a characteristic double-peaked structure. This clear double-peaked feature emerges when either (1) the isotropic explosion energy is small (low-luminosity GRB) or (2) the viewing angle is large (off-axis GRB).
Our results demonstrate that follow-up radio observations carried out within a few years of the optical discovery of nearby Type IcBL SNe (~100 Mpc) can provide a unique diagnostic of off-axis GRBs that would otherwise remain undetected in Type IcBL SNe. This represents a step toward revealing the nature of long GRB progenitors and clarifying their connection to Type IcBL SNe.
\end{abstract}



\section{Introduction} \label{sec:intro}

Long gamma-ray bursts (long GRBs) and broad-lined, hydrogen/helium-deficient supernovae (Type IcBL SNe) are now widely understood to originate from the deaths of massive, stripped-envelope stars. This connection was first established by the discovery of GRB 980425 and its associated SN 1998bw \citep{1998Natur.395..663K,1999AandAS..138..465G} and has since been reinforced by several nearby associations, building a robust empirical link over the past quarter century \citep[e.g.,][]{1998ApJ...504L..87W,2006ARAandA..44..507W,2017AdAst2017E...5C,2024arXiv241108866F}.
Despite the growing number of observed Type IcBL SNe, the fraction of these events that are intrinsically associated with GRBs remains uncertain \citep{2024arXiv241108866F}. The number of GRB–SNe with clearly established associations is reported to be much smaller than the total number of detected Type IcBL SNe. This discrepancy can be attributed to several factors: (1) only a subset of Type IcBL SNe are intrinsically linked to GRB events \citep{2006ApJ...638..930S}; (2) GRB–SNe are observed only when the relativistic jet successfully breaks out of the progenitor surface, while in other cases the jet is choked within the star \citep{2001ApJ...550..410M,2002MNRAS.337.1349R}; and (3) many GRBs associated with Type IcBL SNe are missed because their jets are oriented away from our line of sight (off-axis GRBs; but see also \citealt{2020A&A...639L..11I}).
In the third scenario in particular, off-axis GRBs are expected to be much fainter than on-axis GRBs at early times \citep{2002ApJ...570L..61G}. Furthermore, the typically small opening angles of GRB jets imply a potentially large population of off-axis GRBs relative to on-axis events \citep{2002MNRAS.332..945R}. These effects make it challenging to constrain the intrinsic rate of GRB–SN associations, in contrast to the comparatively well-determined occurrence rate of Type IcBL SNe \citep{2017PASP..129e4201S}.


Radio observations have long been used to probe the intrinsic association between SNe and GRBs. In core-collapse SNe, radio emission generally arises from the interaction of non-relativistic ejecta with the circumstellar material (CSM) surrounding the progenitor star \citep{1982ApJ...259..302C,2006ApJ...641.1029C,2012ApJ...758...81M,2017hsn..book..875C,2019ApJ...885...41M}.
For GRBs, the afterglow spans a broad range of wavelengths, including radio, and is produced as the relativistic jet decelerates through interaction with the ambient medium. Theoretical studies have shown that this afterglow emission can leave distinctive signatures even for observers whose line of sight lies outside the jet opening angle \citep{2002ApJ...570L..61G}.
Motivated by this, \citet{2010Natur.463..513S} proposed that searches for bright radio emission in Type Ibc SNe powered by relativistic outflows provide an effective way to uncover GRB activity, even when no prompt gamma-ray emission is detected. Building on this idea, \citet{2016ApJ...830...42C} used a sample of SNe discovered by the Palomar Transient Factory and followed up with the Very Large Array to place an upper limit on the fraction of Type IcBL SNe that host an off-axis GRB jet.
Beyond such surveys, detailed radio studies of individual Type IcBL SNe have been carried out to search for GRB jet signatures, though their conclusions differ substantially (e.g., SN2007bg: \citet{2013MNRAS.428.1207S}; SN2007uy: \citet{2011ApJ...726...99V}; PTF11qcj: \citet{2021ApJ...910...16P}).


However, we note that the fact that both a GRB afterglow and an SN shock propagate through the same CSM has received relatively little attention. Population and stacking studies typically treat the two phenomena separately adopting a constant density, ISM-like profile $(\rho \propto r^{0})$ for GRBs and a wind-like profile $(\rho \propto r^{-2})$ for SNe. This practice implicitly, and rarely explicitly, assumes that a GRB and its accompanying IcBL SN occupy different environments. In this work, we explicitly address this gap by developing a unified, predictive framework that tracks both the SN shock and the GRB jet self-consistently through a common CSM and predicts the resulting radio light-curve morphologies.


In this paper, we develop a framework for modeling radio emission in GRB–SNe, explicitly accounting for the expectation that the GRB afterglow and SN shock propagate through a common circumstellar environment. We show that, in a particular region of parameter space, the combination of GRB afterglow and SN radio emission can produce a double-peaked radio light curve.
We further provide a systematic mapping of when and how double-peaked radio light curves emerge, how the ordering of the peaks depends on energetics and geometry, and which parameter combinations most effectively diagnose hidden (off-axis) GRBs among IcBL SNe. This, in turn, enables efficient monitoring strategies and allows late-time radio variability to be translated into robust constraints on the true GRB incidence.
To this end, we construct and apply a two-component, single-environment framework that computes the radio emission from (i) a non-relativistic SN shock and (ii) a relativistic, top-hat GRB jet viewed at an arbitrary observer angle, both expanding into the same wind-stratified CSM. Our parameter survey spans isotropic-equivalent energy $E_\mathrm{iso}$, viewing geometry $\theta_\mathrm{obs}/\theta_\mathrm{jet}$, and wind density normalization $A$, yielding both single- and double-peaked behaviors.
These results motivate sustained, multi-year radio monitoring of nearby ($\lesssim 100$ Mpc) IcBL SNe with facilities such as the VLA, MeerKAT, ASKAP, and SKA pathfinders to uncover the off-axis GRB population.


The principal contributions of this paper are threefold: (i) a unified, single-CSM model that self-consistently couples IcBL SN radio emission with GRB afterglows and predicts their composite light-curve morphologies; (ii) an application to SN 2007bg that illustrates how late-time radio peaks can reveal hidden, strongly off-axis GRBs; and (iii) analytic diagnostics and a morphology atlas that delineate the conditions for single versus double peaks and their ordering.
Section \ref{sec:motivation_CSMslope} summarizes the environmental assumptions of previous GRB events (including wind vs. ISM profiles). 
Section \ref{sec:mm} describes our treatment of each component, focusing on the consistency caveats in the calculation of the SN and GRB components. 
Section \ref{sec:07bg} applies the framework to SN 2007bg, and Section \ref{sec:result} presents the parameter survey and morphology classification.
Section \ref{sec:discussion} discuss observational strategies, limitations, and conclusions.

\begin{figure}[htb]
\centering
  \includegraphics[width=0.45\textwidth]{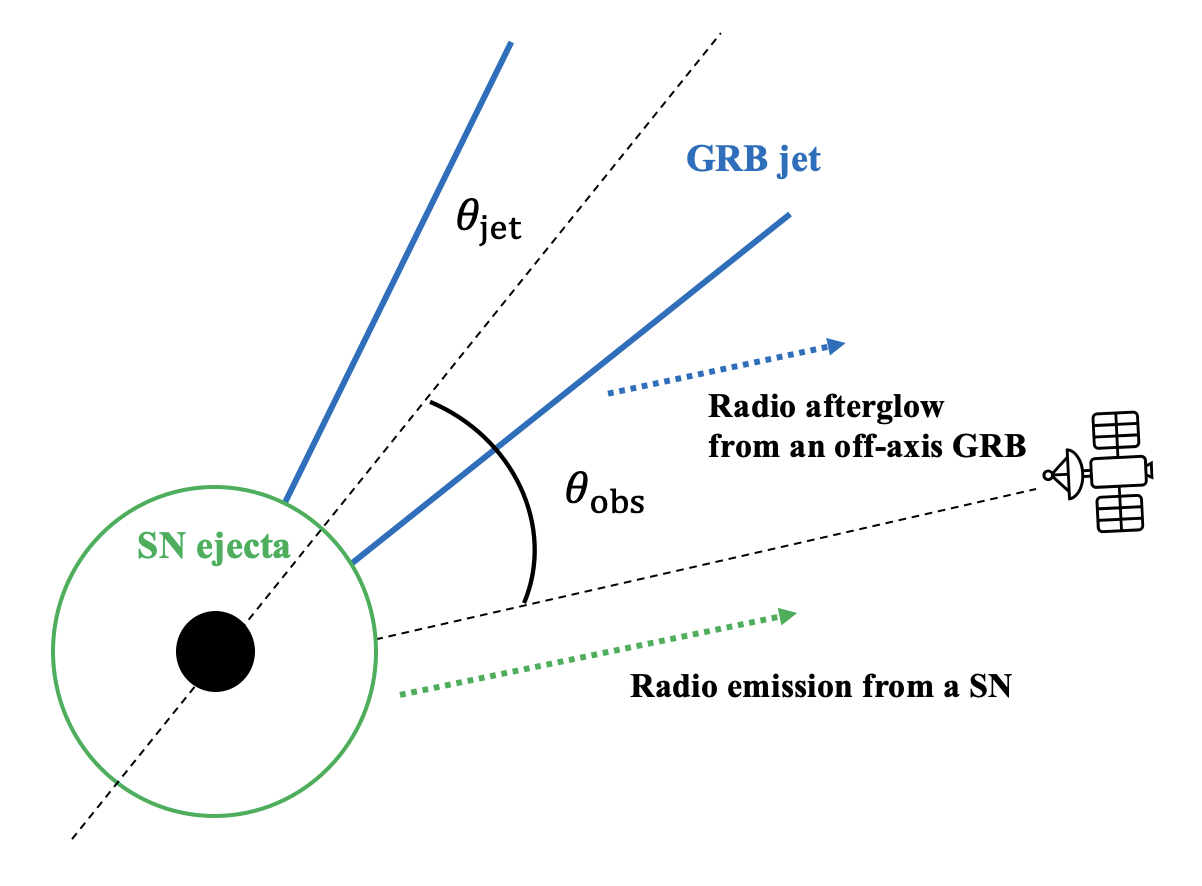}
    \caption{
    Schematic picture of the system assumed in this study. 
    }
\label{fig:image}
\end{figure}



{ \section{Review of constraints on the density slope of ambient medium }\label{sec:motivation_CSMslope} }

\begin{table*}
    \centering
    \begin{threeparttable}
        \caption{List of GRB-SNe whose density slope of ambient medium has been constrained.}
        \begin{tabular}{cccc}
        \hline
        \hline
            GRB & SN & value of $s$ & reference \\
        \hline
            GRB~980425 & SN~1998bw & $s=2$ & \citet{2004ApJ...602..886W} \\
            GRB~011121 & underlying & $s=2$ & \citet{2002ApJ...572L..51P,2002ApJ...572L..45B} \\
            GRB~021211 & underlying & $s=0$ & \citet{2003ApJ...586L...5F,2003AandA...406L..33D} \\
            GRB~030329 & SN~2003dh & $s=0$ & \citet{2003Natur.426..138G,2004ApJ...606..381L} \\
            GRB~031203 & SN~2003lw & $s=2$ & \citet{2005ApJ...625L..91R,2004AandA...419L..21T} \\
            GRB~050525A & SN~2005nc & $s=0,2$ & \citet{2012MNRAS.427..288R} \\
            GRB~060218 & SN~2006aj & $s=2$ & \citet{2006Natur.442.1008C} \\
            GRB~081007 & SN~2008hw & $s=0$ & \citet{2013ApJ...774..114J} \\
            GRB~091127 & SN~2009nz & $s=0$ & \citet{2011AandA...535A..57F, 2010ApJ...718L.150C, 2011ApJ...743..204B, 2011AandA...535A.127V} \\
            GRB~100316D & SN~2010bh & $s=2$ & \citet{2013ApJ...778...18M} \\
            GRB~101219B & SN~2010ma & $s=0$ & \citet{2011ApJ...735L..24S, 2015ApJ...800L..34L} \\
            GRB~111209A & SN~2011kl? & $s=0$ & \citet{2015Natur.523..189G, 2018AandA...617A.122K} \\
            GRB~120422A & SN~2012bz & $s=2$ & \citet{2012AandA...547A..82M, 2012ApJ...756..190Z} \\
            GRB~120714B & SN~2012eb & $s=0,2$ & \citet{2012CBET.3200....1K,2017PhRvD..96j3004T,2019AandA...622A.138K} \\
            GRB~130215A & SN~2013ez & - & \citet{2014AandA...568A..19C} \\
            GRB~130427A & SN~2013cq & $s=0, 2$ & \citet{2013ApJ...776..119L, 2014ApJ...781...37P, 2014Sci...343...48M} \\
            GRB~130702A & SN~2013dx & $s=2$  & \citet{2013ApJ...776L..34S, 2015AandA...577A.116D}  \\
            GRB~130831A & SN~2013fu & $s=0$ & \citet{2016ApJ...823..156Z, 2019AandA...622A.138K} \\
            GRB~140606B & iPTF14bfu & $s=0,2$ & \citet{2015MNRAS.452.1535C} \\
            GRB~161219B & SN~2016jca & $s=0$ & \citet{2018MNRAS.477..153A} \\
            GRB~171010A & SN~2017htp & $s\sim3$ & \citet{2019MNRAS.490.5366M,2019MNRAS.486.2721B} \\
            GRB~171205A & SN~2017iuk & $s=0$ & \citet{2024ApJ...962..117L} \\
            GRB~180728A & SN~2018fip & - & \citet{2019ApJ...874...39W} \\
            GRB~190114C & SN~2019jrj & $s=2$ & \citet{2020ApJ...890....9A, 2022AandA...659A..39M} \\
            GRB~190829A & SN~2019oyw & $s=0$ & \citet{2021Sci...372.1081H, 2024ApJ...977..256B} \\
            GRB~221009A & SN~2022xiw & $s=0 \rightarrow 2$ & \citet{2023ApJ...949L..39S} \\
            GRB~230812B & SN~2023pel & $s=0$ & \citet{2024MNRAS.530....1H} \\
            \hline
            \hline
        \end{tabular}
    \end{threeparttable}
    \label{tab:GRBSN_list}    
\end{table*}

Constraining the nature of GRB–SN progenitors, including the possible presence of an off-axis GRB jet, requires a detailed understanding of their afterglow light-curve behavior, for which a key parameter is the density slope of the ambient medium. Assuming an ambient number density profile $n \propto r^{-s}$, a power-law index of $s = 0$ corresponds to a GRB jet propagating in a uniform ISM, while $s = 2$ describes expansion into a stellar wind emitted by the progenitor star. Table~\ref{tab:GRBSN_list} lists GRB–SNe for which the density slope of the ambient medium has been discussed or constrained in previous studies, compiled from the \texttt{GRBSN Webtool} \citep{2025A&C....5200954F}. As shown there, the value of $s$ is neither uniformly constrained nor consistently adopted across the GRB afterglow literature.

When modeling the propagation of a GRB jet, it is reasonable to assume that the structure of the ambient medium is set by the mass-loss history of the progenitor. If the GRB progenitor is not expected to undergo significant stellar mass loss, for example, in the case of a binary neutron star system thought to power short GRBs, then an ISM-like profile with $s = 0$ can be a reasonable assumption. In contrast, for a GRB accompanied by an SN, the ambient medium should be shaped by the stellar wind from the progenitor, naturally motivating the adoption of $s = 2$ in the literature. This expectation is supported by \citet{2025arXiv250506609M}, who used Markov chain Monte Carlo analysis of radio SNe and found a preference for $s \simeq 2$ in a large fraction of SNe IcBL. In this sense, we argue that adopting $s = 2$ for GRB–SNe should not be dismissed.
Several previous studies have placed upper limits on the fraction of off-axis GRBs hosted by SNe IcBL \citep[e.g.,][]{2014MNRAS.440..821B,2014ApJ...782...42C,2016ApJ...830...42C,2023ApJ...953..179C,2019ApJ...879...89M}, but these works often presuppose that the GRB jet interacts with an ISM-like ambient medium. Consequently, a comprehensive exploration of how the expected light curves depend on model parameters, across a wide range of configurations for both GRB afterglows and radio SNe, remains lacking.

From a theoretical standpoint on the circumstellar environments of GRB–SN progenitors, \citet{2006AandA...460..105V} examined several scenarios that might produce an ISM-like ambient medium around long GRB progenitors, including low metallicity, weak stellar winds, high interstellar medium density and pressure, the presence of a binary companion, and stellar proper motion. These scenarios were explored with the goal of placing the wind termination shock as close to the progenitor as $\sim 0.1\ \mathrm{pc}$, but the study concluded that realizing such an ISM-like environment is possible only within rather limited regions of parameter space.
It is also worth noting that the timescale on which we can observe a GRB jet interacting with an ISM-like ambient medium depends on whether the jet is viewed on-axis or off-axis. For an on-axis GRB jet, the interaction with the ISM-like region becomes observable after $ t_{\rm obs} \sim R_{\rm ISM}/(\Gamma^2 c)$, where $R_{\rm ISM}$ is the inner radius of the ISM-like ambient medium and $\Gamma$ is the Lorentz factor of the GRB jet (as introduced in the following section). This timescale is shorter by a factor of $\Gamma^2$ than in the non-relativistic case.
However, if the GRB jet is viewed off-axis, the situation changes: the relevant timescale becomes simply $t_{\rm obs} \sim R_{\rm ISM}/c$. This implies that the naive assumption of $s = 0$ immediately after the onset of the afterglow can be misleading and may obscure a physically plausible interpretation of the ambient medium properties.

It should also be emphasized that the value of $s$ inferred from observational fitting cannot always be interpreted straightforwardly. Photometric and spectroscopic fitting typically yields $s$ together with $p$, the spectral index of the accelerated electrons. However, \citet{2013ApJ...776L..34S} demonstrated that using only optical band data does not provide sufficient leverage to clearly distinguish between ISM-like and wind-like ambient medium models, limiting the robustness of constraints on $s$.

Given these arguments, we adopt the view that the power-law index describing the ambient medium around GRB–SN progenitors is not yet comprehensively understood. To achieve a unified picture of this parameter, it is essential to develop a more complete understanding of GRB afterglow behavior. In particular, we expect that afterglows from GRB–SNe are governed by the combined contributions of SN radio emission and GRB afterglow emission.
In this paper, we present a comprehensive demonstration of the expected observational signatures from GRB–SNe, based on a framework in which both the GRB jet and the SN shock interact with a wind-like ambient medium. Following the method described in Section~\ref{sec:mm}, we explore the possible diversity of radio emission properties in GRB–SNe and illustrate this with a practical example: model fitting to SN~2007bg, whose radio light curve exhibits a characteristic double-peaked structure.

\section{Model and Method} \label{sec:mm}

In this study, we consider two shock waves of SN and GRB ejecta propagating in the same environment characterized by stellar wind of a progenitor star.
We consider the density structure of CSM as
\begin{equation}
  n_\mathrm{CSM}(r)=3\times10^{35}A r^{-2}\ {\rm cm}^{-3},
\end{equation}
where $A=(\dot{M}/10^{-5}\ M_\sun{\rm yr}^{-1})(v_{\rm wind}/10^3\ {\rm km}\ {\rm s}^{-1})^{-1}$ reflects the wind mass-loss rate of a typical Wolf-Rayet star \citep{2012ApJ...758...81M,2018MNRAS.478..110S}. 

Around shock waves, electrons are accelerated through diffusive shock acceleration \citep[][]{1978ApJ...221L..29B,1978MNRAS.182..147B}, magnetic reconnection \citep{2011ApJ...726...75S,2015MNRAS.450..183S}, or turbulent acceleration \citep{2009ApJ...705.1714A,2014ApJ...780...64A}. 
The injection rate of comoving spectral number density is described as a single power of the Lorentz factor $\gamma$ of electrons: $Q(\gamma)\propto\gamma^{-p},\ \ (\gamma_{\rm m}\leqq\gamma\leqq\gamma_{\rm M})$ \citep{1998ApJ...497L..17S}.
The theoretical prediction of the spectral index is $p=2$ for non-relativistic shocks \citep{1978ApJ...221L..29B,1978MNRAS.182..147B} or $p\sim2.2$ for relativistic shocks \citep{2005PhRvL..94k1102K}. However, values inferred from observed radio SNe and GRB afterglows typically suggest $2 < p < 3$ \citep{2006ApJ...641.1029C,2020ApJ...896..166R}. 
We leave the difference in the spectral index $p$ between the SNe and the GRB afterglow model.

The minimum Lorentz factor $\gamma_{\rm m}$ is computed as 
\begin{equation}
    \gamma_{\rm m}=
    \left\{
    \begin{array}{ll}
        \epsilon_{\rm e}\frac{p-2}{p-1}\frac{m_{\rm p}}{m_{\rm e}}(\Gamma_{\rm }-1)\ \ \ \ \ (\rm for\ GRB)\\
        2\ \ \ \ \ \ \ \ \ \ \ \ \ \ \ \ \ \ \ \ \ \ \ \ \ \ \ \ \ (\rm for\ SN)
    \end{array}
    \right.,
    \label{eq:gamma_m}
\end{equation}
where $m_{\rm p}$ and $m_{\rm e}$ are the proton and electron mass. We introduced a microphysics parameter $\epsilon_{\rm e}$ \citep[e.g.][]{1998ApJ...497L..17S,2020ApJ...896..166R}. For GRB case, we require the Lorentz factor of the bulk of the shocked CSM $\Gamma$. For SN, we assume $\gamma_{\rm m}\sim2$ in this study because the velocity of the SN shock is in a non-relativistic regime (\citet[][]{2023MNRAS.524.6004W}, but see also \citet[][]{2024ApJ...960...70M}). 
The maximum Lorentz factor is given by the balance of acceleration with the Bohm limit and synchrotron cooling: $\gamma_{\rm M}=\sqrt{\frac{6\pi e}{\sigma_{\rm T}B}}$ \citep{2006MNRAS.369..197F}, where $\sigma_{\rm T}$ is the Thomson scattering cross section and $e$ is the elementary charge. A fraction $\epsilon_{\rm B}$ of the internal energy density $\varepsilon_{\rm sh}=4\Gamma_{\rm }(\Gamma_{\rm }-1)n_{\rm CSM}m_{\rm p}c^2$ can be transferred to the turbulent magnetic field $B=\sqrt{8\pi \epsilon_{\rm B}\varepsilon_{\rm sh} }$ \citep{1998ApJ...497L..17S}. 

The spectra of these non-thermal particles are modified by radiative cooling. 
The steady state particle distribution can be written as \citep{2024ApJ...960...70M}
\begin{eqnarray}
    n(\gamma)=Ct_{\rm cool}(\gamma)\gamma^{-1} \max(\gamma,\gamma_m)^{1-p},
    \label{eq:PED}
\end{eqnarray}
where $t_{\rm cool}(\gamma)$ is the total cooling time scale including synchrotron, inverse Compton, and adiabatic cooling \citep{1998ApJ...509..861F, 2019ApJ...885...41M, 2021ApJ...918...34M, 2022ApJ...936...98B}. 
The cooling break $\gamma_{\rm c}$ is determined by the balance of the radiative cooling time scale and the dynamical time scale \citep{2024ApJ...960...70M,2025MNRAS.536.1822K}. 
The normalization factor $C$ can be derived by assuming the balance between injection and cooling of relativistic electrons at any time \citep{1998ApJ...509..861F,2024ApJ...960...70M}
\begin{eqnarray}
    C = 
    \left\{
    \begin{array}{ll}
        \frac{4(p-1)\Gamma_{\rm }n_{\rm CSM} \Gamma_{\rm }c}{\gamma_{\rm m}^{1-p} R_{\rm sh}}\ \ \ \ (\rm for\ GRB)\\
        \frac{9(p-2)\epsilon_{\rm e}n_{\rm CSM}m_{\rm p} V_{\rm sh}^3}{8\gamma_{\rm m}^{2-p}m_{\rm e} c^2 \Delta R}\ \ (\rm for\ SN)
        \label{eq:Ngamma}
    \end{array}
    \right.,
\end{eqnarray}
where $V_{\rm sh}$ is the velocity of the shock surface. 


We calculate synchrotron radiation and synchrotron self-absorption (SSA) using fitting formulae for the synchrotron function \citep{2010PhRvD..82d3002A,2013RAA....13..680F} under thin shell approximation. The shell width of the shocked wind {measured in its rest frame} is estimated as $\Delta R \equiv R_{\rm sh}/(4 \Gamma_{\rm })$ for GRBs. Meanwhile, $\Delta R \equiv f R_{\rm sh}$ is adopted for SNe, where $f$ is the geometric thickness factor depending on the choice of $n$ \citep{1982ApJ...258..790C,1994ApJ...420..268C,1999ApJS..120..299T} (usually $f\sim0.2$).
The optical depth of SSA is given by $\tau_{\rm SSA}=\alpha_{\rm SSA}\Delta R_{\rm }$, where $\alpha_{\rm SSA}$ is the absorption coefficient for SSA \citep{1986rpa..book.....R}. The observed radio luminosity is suppressed by a factor of $(1-\exp(-\tau_{\rm SSA}))/\tau_{\rm SSA}$ \citep[e.g.][]{2024ApJ...960...70M,2025MNRAS.536.1822K}.

For GRB, we consider the Equal Arrival Time Surface (EATS) for the observed radiation flux \citep{1997ApJ...491L..19W,2005ApJ...631.1022G}. To calculate the observed radiation flux from off-axis afterglows, we should integrate over the solid angle along with EATS \citep{1999ApJ...523..187W,2024MNRAS.528.2066P}. 
We take $\theta=0$ as the line of sight. 
We do not consider the EATS effects for SNe because the speed of shock wave is non-relativistic.

\subsection{The dynamics of GRB afterglows}\label{subsec:GRB_dynamics}

In this subsection, we describe the interaction between the wind and the cold top-hat jet with the initial Lorentz factor $\Gamma_0$, isotropic equivalent energy $E_{\rm iso}$, and the opening angle $\theta_{\rm jet}$. 
In general, the width of the ejecta and magnetization may modify the early phase evolution \citep{2000ApJ...542..819K,2009AandA...494..879M,2014MNRAS.442.3495V,2023MNRAS.526..512K,2025mnras/staf879K,2025/mnras/staf1923K,2025ApJ...990..110W}. Since our interest in this study is the late afterglow, we simply neglected these effects.

We use the semi-analytic model given in \citet{2018ApJ...865...94D}, where we adopt $P_k=2.0$ and $Q_k=1.6$ in Eqs. (28), (29), (30), and (31) in their paper.
In their model, the dynamics of the shock wave is characterized by 4 phases.
At first, the dynamics of the shock in their model is a free expansion with constant $\Gamma_0$ until the deceleration radius $R_{\rm dec}=E_{\rm iso}/(4\pi n_{\rm CSM}r^2m_{\rm p}c^2\Gamma_0^2)$ \citep{1995ApJ...455L.143S}. 
Beyond $R_{\rm dec}$, the shock decelerates adiabatically following the Blandford-McKee (BM) solution $\Gamma_{\rm } \propto t^{-1/2}$ \citep{1976PhFl...19.1130B,2008AandA...478..747G,2013NewAR..57..141G}. 
The dynamics of the shock wave is further modified after the causal connection around $\Gamma_{\rm }\sim\theta_{\rm jet}^{-1}$, starting lateral spreading \citep{1999ApJ...525..737R,2012ApJ...751..155V,2018ApJ...865...94D,2020ApJ...896..166R,2024ApJS..273...17W}. Finally, the shock decelerates non-relativistic following the Sedov-Taylor solution $\Gamma_{\rm }\beta_{\rm } \propto t^{-1/3}$ \citep{1988RvMP...60....1O}.

\subsection{The dynamics of radio SNe}

In this subsection, we describe the interaction between the wind and SN ejecta with the SN ejecta mass $M_{\rm ej}$, the kinetic energy $E_{\rm ej}$, and its density structure consisting of a flat inner component and a steep outer layer with a density slope $n=7$ \citep{1982ApJ...259..302C,1982ApJ...258..790C}. 
We follow so-called thin-shell approximation where the thin shell consisting of the shocked CSM and the shocked ejecta is pushed outward by the ram pressure of the ejecta \citep{1982ApJ...259..302C,2017hsn..book..875C}. Assuming that the location of the shock is almost the same as that of the thin shell, we can describe the explicit formula of the time evolution of the shock wave \citep[for the concrete description, see Eqs. (4), (5), (6) in][]{2024ApJ...960...70M}.
We note that the ejecta mass and kinetic energy control the radius and velocity of the shock wave, and the light curve peak in radio SNe is characterized by SSA \citep[e.g.,][]{1998ApJ...499..810C}.

{ \section{Candidates of Off-axis GRB-SN}  \label{sec:07bg} }

In this section, we present an example application of our framework to demonstrate the validity of our approach. A suitable target is a Type IcBL SN that exhibits a peculiar radio light-curve evolution indicative of an off-axis GRB.
Although numerous theoretical studies have explored the possible contributions of a magnetar \citep{2025MNRAS.539.1908O} and cocoon emission \citep{2018ApJ...863...32D,2022MNRAS.512.3627D} to the radio light curves of GRB–SNe, only a few candidates for SNe accompanied by off-axis GRBs have been identified. XRF 020903 is one such candidate: it displays a double-peaked radio light curve, with peaks at $\mathcal{O}(1)$ days (GRB component; \citealt{2015ApJ...806..222U,2021ApJ...915...46C}) and $\mathcal{O}(10)$ days (SN component; \citealt{2006ApJ...643..284B}). Because the viewing angle is slightly off-axis, similar to GRB 170817A \citep{2018MNRAS.478L..18T,2019MNRAS.489.1919T}, this object is classified as an X-ray flash, a sub-class of GRBs.

Another candidate is SN 2007bg, one of the brightest radio-SNe classified as a Type IcBL SN whose radio light curves have double peaks at $\mathcal{O}(100)$ days and $\mathcal{O}(1000)$ days \citep{2013MNRAS.428.1207S}. 
The host galaxy of SN 2007bg was a low-metallicity galaxy similar to GRB-related SNe hosts based on optical photometry and spectroscopy observations \citep{2010AandA...512A..70Y}.
A stratified CSM scenario \citep{2013MNRAS.428.1207S} and an extreme particle acceleration scenario \citep{2024ApJ...960...70M} are proposed to explain the double-peaked radio light curves of this SN, but the plausibility of a possible off-axis GRB association has not been thoroughly explored.

\begin{figure*}
    \centering
    \includegraphics[width=\linewidth]{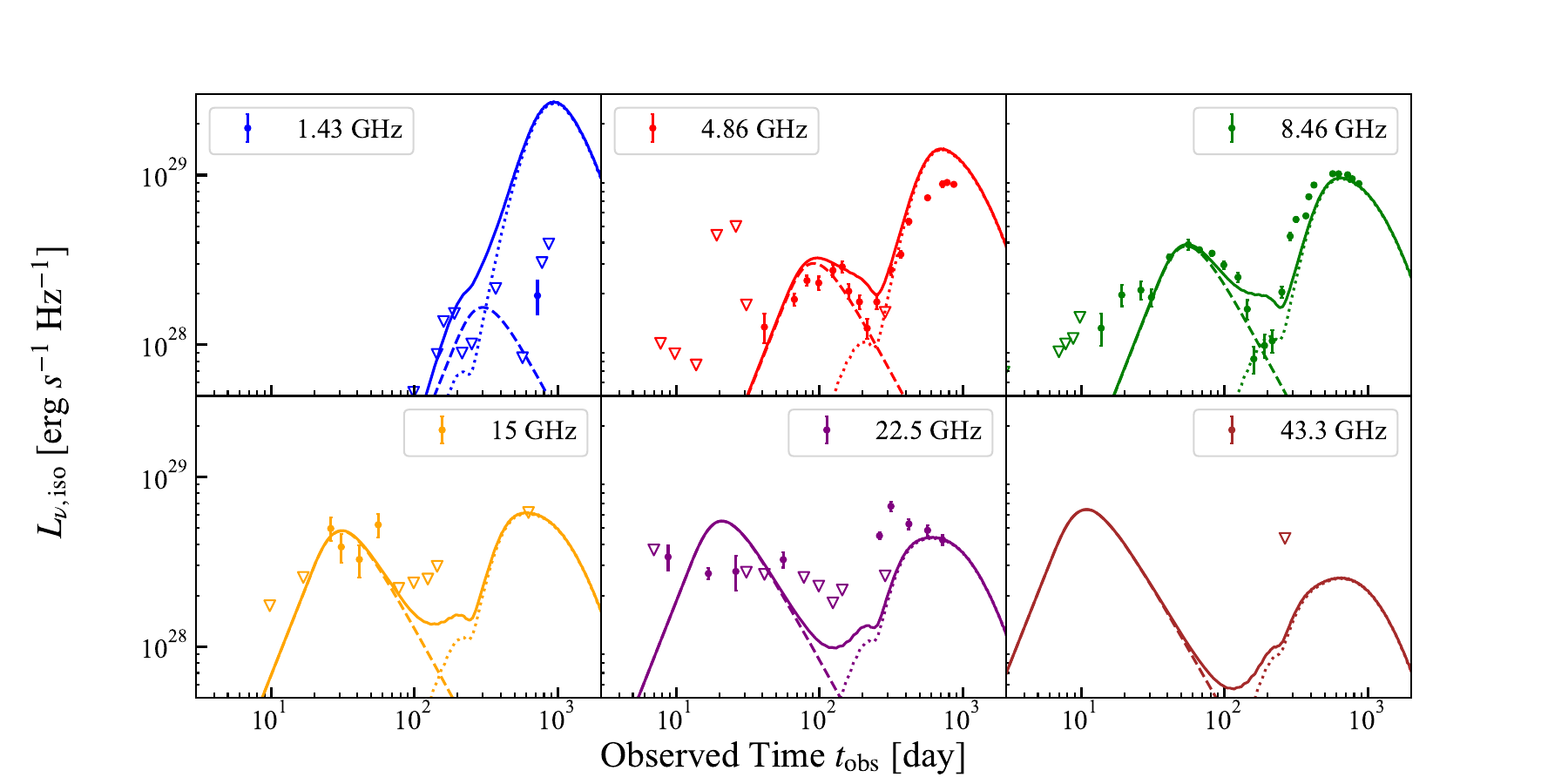}
    \caption{Multi-band radio light curves for SN 2007bg. Each panel corresponds to the different radio frequency. The filled circles are observed data, and open triangles are upper limit. Our demonstrated light curves are shown in dashed (SN), dotted (GRB), and solid (total) lines. The parameters are $E_0=3\times10^{54}$ erg, $A=10\ \rm cm^{-1}$, $p_{\rm GRB}=2.5$, $\epsilon_{e,\rm GRB}=10^{-1}$, $\epsilon_{B,\rm GRB}=10^{-2}$, $\theta_{\rm jet}=0.05$ rad, $\theta_{\rm obs}=1.2$ rad. For SN, $p_{\rm SN}=3.0$, $\epsilon_{e,\rm SN}=3\times10^{-1}$ and $\epsilon_{B,\rm SN}=5\times10^{-2}$.
    \label{fig:GRB_test}}
\end{figure*}

We apply our model to explain the double peaks of the radio light curves of SN 2007bg. 
We use $M_{\rm ej}=1.5M_\odot$ and $E_{\rm ej}=4\times10^{51}$ erg reported in \citet[][]{2010AandA...512A..70Y}. 
Our demonstration results are shown in Figure \ref{fig:GRB_test} and Table \ref{table:models}.
The first peak corresponds to the radio SN model and the second peak corresponds to the off-axis afterglow model. The peaks are achieved when the optical depth of SSA emission is below 1 for SN and when the whole jet is visible for GRB. A large isotropic-equivalent energy $E_{\rm iso}$ and the angle ratio $\theta_{\rm obs}/\theta_{\rm jet}$ are required to reproduce the radio peak at $\mathcal{O}(1000)$ days. The intrinsic energy of the afterglow is estimated as $E_{\rm int}=E_{\rm iso}(\theta_{\rm jet}^2/2)\approx10^{51}$ erg, which is comparable to the explosion energy of the SN 2007bg \citep{2013MNRAS.428.1207S}. 
The inferred wind parameter $A=10$ is about an order of magnitude larger than the typical value of Wolf-Rayet stars observed in the Milky Way \citep{2000AandA...360..227N}, but such a large value of $A$ may be naturally expected by accounting for a pre-SN neutrino emission \citep{2025ApJ...982...93S}.

The overall agreement between the model and the data is moderate.
The discrepancy at $\sim 20$ day in the 8.46 GHz and $\sim10$ day in the 22.5 GHz light curves, where the model underestimates the observed data, suggests that there might be late time energy injection or additional component from reverse shock, which is sometimes discussed in radio flare of GRB afterglows \citep[e.g.][]{2014MNRAS.440.2059A,2023MNRAS.523.4992A,2024ApJ...970..139S}.
In addition, our model slightly overestimates the data at $\sim700$ day in the 4.86 GHz and $\sim200$ day in the 8.46 GHz light curve. 
This may be due to the simplicity of our model, in which both radio afterglows and radio SNe are simply summed up. 
However, in general, the spatial distribution and evolution of both components should be considered simultaneously. 
Nevertheless, these slight deviations are within a factor of $\sim 2$.
Although a large discrepancy still exists in the 1.43 GHz, 
this might mean that the additional contribution of the supra-thermal electrons to SSA is necessary \citep{2017ApJ...845..150R,2018MNRAS.480.4060W}.
Our unified CSM model successfully reproduces the overall trend of the light curve, including the observed double-peaked structures.

As demonstrated in this section, an off-axis afterglow can contribute to the radio light curves of a SN IcBL. The double-peaked structure of the radio light curves can be one of the evidences of the off-axis GRB association. The time scale for the contribution of radio afterglow is $O(1000)$ days (but can be diverse as described in the following Section \ref{sec:result}).
Therefore, it is important to carry out follow-up observations for radio SN at least a few years after the explosion. We suggest that future radio telescopes such as SKA and Busting Universe Radio Survey Telescope in Taiwan (BURSTT) will increase the number of off-axis GRB associated SNe. 

\begin{table*}
 \caption{Parameters used in the model demonstration. 
 From the left, isotropic-equivalent energy $E_{\rm iso}$, bulk Lorentz factor $\Gamma_0$, wind density parameter $A$, spectral index of accelerated electrons $p_{\rm GRB}$, electron energy ratio $\epsilon_{e,\rm GRB}$, and magnetic energy ratio $\epsilon_{B,\rm GRB}$.}
 \label{table:models}
 \centering
  \begin{tabular}{cccccccc||cccccc}
   \hline
   $E_{\rm iso}$ [erg] & $\Gamma_0$ & $\theta_{\rm jet}$ [rad] & $\theta_{\rm obs}$ [rad] & $A$ $[{\rm cm}^{-1}]$ & $p_{\rm GRB}$ & $\epsilon_{e,\rm GRB}$ & $\epsilon_{B,\rm GRB}$ & $M_{\rm ej}$ [$M_\odot$] & $E_{\rm ej}$ [erg] & $p_{\rm SN}$ & $\epsilon_{e,\rm SN}$ & $\epsilon_{B,\rm SN}$ \\
   \hline \hline
   $3\times10^{54}$ & $100$ & $0.05$ & $1.2$ & $10$ & $2.5$ & $10^{-1}$ & $10^{-2}$ & $1.5$ & $4\times10^{51}$ & $3.0$ & $3\times10^{-1}$ & $5\times10^{-2}$ \\
   \hline
  \end{tabular}
\end{table*}

\section{Peak time and peak flux}

Since our model considers the unified CSM scenario for both radio SN and off-axis radio afterglow, it is worthwhile to calculate the ratio of the peak time and peak flux of both components, which might be useful to identify the characteristic double-peak behavior in the radio light curves.

\subsection{The case of radio SN}

It would be useful to explicitly show the parameter dependence of the peak time and luminosity in radio SNe. Here we just assume that adiabatic cooling dominates the cooling process of relativistic electrons accelerated by the SN shock, and that the spectral index of the electron energy distribution is fixed to be $p=3$. Based on the fact that the peak is characterized by SSA \citep{1998ApJ...499..810C}, both of these can be written down as follows:
\begin{eqnarray}
    t_{\rm p,SN} &\sim& 5\,{\rm days} \left(\frac{\epsilon_{e,\rm SN}}{10^{-1}}\right)^{\frac{2}{7}}
    \left(\frac{\epsilon_{B,\rm SN}}{10^{-2}}\right)^{\frac{5}{14}}
    \left(\frac{A}{1}\right)^{\frac{9}{14}}\nonumber \\
    &\times&\left(\frac{V_{\rm sh}}{10^9\,{\rm cm\,s}^{-1}}\right)^{\frac{2}{7}}
    \left(\frac{\nu}{10\,{\rm GHz}}\right)^{-1}, \\
    F_{\rm p,SN} &\sim&  10^{2}\,\mu{\rm Jy}\ \left( \frac{D_{L}}{100\ {\rm Mpc}} \right)^{-2} \left(\frac{\epsilon_{e,\rm SN}}{10^{-1}}\right)^{\frac{5}{7}}
    \left(\frac{\epsilon_{B,\rm SN}}{10^{-2}}\right)^{\frac{9}{14}}\nonumber \\
    &\times&\left(\frac{A}{1}\right)^{\frac{19}{14}}
    \left(\frac{V_{\rm sh}}{10^9\,{\rm cm\,s}^{-1}}\right)^{\frac{19}{17}},
\end{eqnarray}
where the weak temporal dependence of the SN shock velocity is neglected for simplicity. Note the independence of $F_{\rm p,SN}$ on the frequency\footnote{If we include the time dependence of $V_{\rm sh}$ on $t_{\rm p,SN}$ through the weak temporal dependence of the self-similar solution \citep{1982ApJ...258..790C}, there appears a weak dependence of $F_{\rm p,SN}$ on the frequency}.

\subsection{The case of off-axis radio afterglow}

For an observer viewing a uniform jet with angle $\theta_{\rm obs}$ from the center of the jet, the observed emission differs from that of an observer viewing the jet from the on-axis. An "off-axis" means that viewing angle is larger than the opening angle of the jet ($\theta_{\rm obs}>\theta_{\rm jet}$) so that the Doppler factor is reduced to be
\begin{equation}
\begin{split}
    \mathcal{D}(\theta)&=\frac{1}{\Gamma(1-\beta\cos(\theta_{\rm obs}-\theta_{\rm jet}))}\\
    &\simeq\frac{2\Gamma}{1+\Gamma^2(\theta_{\rm obs}-\theta_{\rm jet})^2}.
\end{split}
\end{equation}
The observed time scale is also prolonged as (neglecting redshift correction for simplicity)
\begin{equation}
\begin{split}
    t_{\rm obs}(\theta_{\rm obs})&=t-\frac{R}{c}\cos(\theta_{\rm obs}-\theta_{\rm jet}) \\
    &\simeq\frac{R}{2\Gamma^2c}( 1+\Gamma^2(\theta_{\rm obs}-\theta_{\rm jet})^2).
\end{split}
\end{equation}
Thus, the observed flux is scaled as \citep[e.g.][]{2002ApJ...570L..61G}
\begin{equation}
    F_\nu(\theta_{\rm obs},t_{\rm obs})=a^3F_{\nu/a}(0,at_{\rm obs}),
    \label{eq:off_L}
\end{equation}
where $a=[1+\Gamma^2(\theta_{\rm obs}-\theta_{\rm jet})^2]^{-1}$. 
If the emission region is far from an observer in the direction $\theta_{\rm obs}$, or in turn $\Gamma(\theta_{\rm obs}-\theta_{\rm jet})>1$, then the emission is initially dim for the observer. As the jet decelerates, the angle of the relativistic beaming cone $\Gamma^{-1}$ increases.
Then, the off-axis afterglow can be visible when $\Gamma(\theta_{\rm obs}-\theta_{\rm jet})=1$ \citep[see e.g.][in their Appendix]{2024JHEAp..41....1O}.
As the jet continues to decelerate, the core of the jet is visible when $\Gamma\theta_{\rm obs}=1$.
Thus, the peak time of the observed flux for the off-axis observer is given as \citep[under the non-spreading approximation]{2020MNRAS.493.3521B}
\begin{equation}
    t_{\rm peak}=t_{\rm dec}(\Gamma_0\theta_{\rm jet})^{\frac{2(4-s)}{3-s}}\left( \frac{\theta_{\rm obs}}{\theta_{\rm jet}} \right)^{\frac{2(4-s)}{3-s}},
\end{equation}
where
\begin{equation}
    t_{\rm dec}\sim 10^{-1}\ {\rm s}\ \left( \frac{E_{\rm iso}}{10^{52}\ {\rm erg}} \right)\ \left( \frac{A}{1} \right)^{-1}\ \left( \frac{\Gamma}{100} \right)^{-4}
\end{equation}
is a typical deceleration time scale for on-axis observer \citep{2018pgrb.book.....Z}.
For wind-like medium $s=2$, the peak time scale $t_{\rm peak}$ is estimated as
\begin{equation}
    t_{\rm peak}\sim40\ {\rm days}\ \left( \frac{E_{\rm iso}}{10^{52}\ {\rm erg}} \right)\ \left( \frac{A}{1} \right)^{-1}\ \left( \frac{\theta_{\rm obs}}{0.8\ {\rm rad}} \right)^4.
    \label{eq:peak}
\end{equation}
If the peak achieves at $t_{\rm peak}$, the $\nu_{\rm m}$ has already passed. Thus,  
the flux is estimated from Eq. (\ref{eq:off_L}) with $a\simeq2$, considering the spectral shape at $\nu_{\rm obs}>\nu_{\rm m}$ as 
\begin{equation}
\begin{split}
    F_{\rm peak} & =  a^3F_{\nu_{\rm m}/a}(0,at_{\rm peak}) \left( \frac{\nu_{\rm obs}}{\nu_{\rm m}} \right)^{-\frac{p-1}{2}}\\
    &\sim2^{\frac{7-12p}{3}}\times10^{6}\ \mu{\rm Jy}\ \left( \frac{D_{L}}{100\ {\rm Mpc}} \right)^{-2}\ \left( \frac{\nu_{\rm obs}}{10\ {\rm GHz}\ } \right)^{-\frac{3p-5}{6}} \\
    & \times \ \left( \frac{\epsilon_{e,\rm GRB}}{10^{-1}} \right)^{\frac{3p-5}{3}}\ \left( \frac{\epsilon_{B,\rm GRB}}{10^{-2}} \right)^{\frac{3p-2}{12}}\ \left( \frac{A}{1} \right)^{\frac{3p+1}{4}}\\
    & \times \ \left( \frac{E}{10^{52}\ {\rm erg}} \right)^{-\frac{3p-5}{6}} \ \left( \frac{\theta_{\rm obs}}{0.8\ {\rm rad}} \right)^{-3(p-1)}
    \end{split}
\end{equation}

Meanwhile, $\nu_{\rm m}$ also depends on the viewing angle through the Doppler shift, such that for an off-axis observer $\nu_{\rm m}(\theta_{\rm obs}>\theta_{\rm jet})=\nu_{\rm m}(\theta_{\rm obs}=0)/a$. 
However, if the peak achieved at $\nu_{\rm m}$ passing, the entire jet should already be visible, implying $\Gamma\theta_{\rm obs}<1\to a\simeq1$. 
As a result, the peak time due to $\nu_{\rm m}$ passing is given as
\begin{equation}
\begin{split}
    t_{\rm m}\sim 1\ &{\rm day}\ \left( \frac{E_{\rm iso}}{10^{52}\ {\rm erg}} \right)^{\frac{1}{3}}\ \left( \frac{\nu_{\rm obs}}{10\ {\rm GHz}} \right)^{-\frac{2}{3}} \\
    &\times\ \left( \frac{\epsilon_{e,\rm GRB}}{10^{-1}} \right)^{\frac{4}{3}}\ \left( \frac{\epsilon_{B,\rm GRB}}{10^{-2}} \right)^{\frac{1}{3}}.
    \label{eq:tm}
\end{split}
\end{equation}
The peak flux can be estimated from Eq. (\ref{eq:off_L}): 
\begin{equation}
\begin{split}
    F_{\rm m}
    \sim \ & 2\times10^{6}\ \mu{\rm Jy}\ \left( \frac{D_{L}}{100\ {\rm Mpc}} \right)^{-2}\ \left( \frac{E_{\rm iso}}{10^{52}\ {\rm erg}} \right)^{\frac{1}{3}}\ \left( \frac{A}{1} \right)\ \\
    & \times\ \left( \frac{\nu_{\rm obs}}{10\ {\rm GHz}} \right)^{\frac{1}{3}} \left( \frac{\epsilon_{e,\rm GRB}}{10^{-1}} \right)^{-\frac{2}{3}}\ \left( \frac{\epsilon_{B,\rm GRB}}{10^{-2}} \right)^{\frac{1}{3}}\ 
\end{split}
\end{equation}
When $t_{\rm peak}<t_{\rm m}$, the peak of the radio component $\nu_{\rm obs}<\nu_{\rm m}$ is determined by $t_{\rm m}$, which is roughly evaluated by the following condition:
\begin{equation}
\begin{split}
    \frac{\theta_{\rm obs}}{\theta_{\rm jet}} <&\ 6\ 
    \left( \frac{E_{\rm iso}}{10^{52}\ {\rm erg}} \right)^{-1/6}\ \left( \frac{A}{1} \right)^{\frac{1}{4}}\ \left( \frac{\theta_{\rm jet}}{0.05\ {\rm rad}} \right)^{-1}\\
    &\times\ \left( \frac{\epsilon_{e,\rm GRB}}{10^{-1}} \right)^{\frac{1}{3}}\ \left( \frac{\epsilon_{B,\rm GRB}}{10^{-2}} \right)^{\frac{1}{12}}\ \left( \frac{\nu_{\rm obs}}{10\ {\rm GHz}} \right)^{-\frac{1}{6}}
\end{split}
\label{eq:condition}
\end{equation}
On the other hand, the peak is determined by $t_{\rm peak}$ for the case of $t_{\rm peak}>t_{\rm m}$.

\subsection{Ratio of peak time and flux}

When Eq. (\ref{eq:condition}) is satisfied, the peak time of the afterglow is determined by $t_{\rm m}$. 
Thus, we find that the ratio of the peak time is given as
\begin{equation}
\begin{split}
    \frac{t_{\rm m}}{t_{\rm p,SN}}\sim 0.2\ &\left( \frac{E_{\rm iso}}{10^{52}\ {\rm erg}} \right)^{\frac{1}{3}}\ \left(\frac{A}{1}\right)^{-\frac{9}{14}}\ \left(\frac{V_{\rm sh}}{10^9\,{\rm cm\,s}^{-1}}\right)^{-\frac{2}{7}}\\
    &\times\ \left( \frac{\epsilon_{e,\rm GRB}}{10^{-1}} \right)^{\frac{4}{3}}\ \left( \frac{\epsilon_{B,\rm GRB}}{10^{-2}} \right)^{\frac{1}{3}}\ \left( \frac{\nu_{\rm obs}}{10\ {\rm GHz}} \right)^{\frac{1}{3}}\\
    &\times\ \left( \frac{\epsilon_{e,\rm SN}}{10^{-1}} \right)^{-\frac{2}{7}}\ \left( \frac{\epsilon_{B,\rm SN}}{10^{-2}} \right)^{-\frac{5}{14}},
    \label{eq:tm/tp}
\end{split}
\end{equation}
and the flux ratio is
\begin{equation}
\begin{split}
    \frac{F_{\rm m}}{F_{\rm p,SN}} &\sim 10^5\ \left( \frac{E_{\rm iso}}{10^{52}\ {\rm erg}} \right)^{\frac{1}{3}}\ \left( \frac{A}{1} \right)^{-\frac{5}{14}}\ \\
    &\times\ \left( \frac{\nu_{\rm obs}}{10\ {\rm GHz}} \right)^{\frac{1}{3}} \left( \frac{\epsilon_{e,\rm GRB}}{10^{-1}} \right)^{-\frac{2}{3}}\ \left( \frac{\epsilon_{B,\rm GRB}}{10^{-2}} \right)^{\frac{1}{3}}\ \\
    &\times\ \left(\frac{\epsilon_{e,\rm SN}}{10^{-1}}\right)^{-\frac{5}{7}}
    \left(\frac{\epsilon_{B,\rm SN}}{10^{-2}}\right)^{-\frac{9}{14}}\ \left(\frac{V_{\rm sh}}{10^9\,{\rm cm\,s}^{-1}}\right)^{-\frac{19}{17}}.
    \label{eq:fm/fp}
\end{split}
\end{equation}
The both ratio is independent of the viewing angle $\theta_{\rm obs}$.

If Eq. (\ref{eq:condition}) is not satisfied, the peak time of the afterglow is determined by $t_{\rm peak}$. Thus, we find
\begin{equation}
\begin{split}
    \frac{t_{\rm peak}}{t_{\rm p,SN}}\sim 8\ &\left( \frac{E_{\rm iso}}{10^{52}\ {\rm erg}} \right)\ \left( \frac{A}{1} \right)^{-\frac{23}{14}}\ \left( \frac{\theta_{\rm obs}}{0.8\ {\rm rad}} \right)^4\\
    &\times\left(\frac{\epsilon_{e,\rm SN}}{10^{-1}}\right)^{-2/7}
    \left(\frac{\epsilon_{B,\rm SN}}{10^{-2}}\right)^{-5/14}
    \left(\frac{\nu_{\rm obs}}{10\,{\rm GHz}}\right)\\
    &\times\left(\frac{V_{\rm sh}}{10^9\,{\rm cm\,s}^{-1}}\right)^{-2/7},
    \label{eq:tpeak/tp}
\end{split}
\end{equation}
and the flux ratio is
\begin{equation}
\begin{split}
    \frac{F_{\rm peak}}{F_{\rm p,SN}} & \sim2^{\frac{4(1-3p)}{3}}\times 10^{4}\ \left( \frac{\nu_{\rm obs}}{10\ {\rm GHz}\ } \right)^{-\frac{3p-5}{6}} \\
    & \times \left( \frac{E}{10^{52}\ {\rm erg}} \right)^{-\frac{3p-5}{6}}\ \left( \frac{A}{1} \right)^{\frac{42p-43}{56}}\ \left( \frac{\theta_{\rm obs}}{0.8\ {\rm rad}} \right)^{-3(p-1)}\\
    & \times \ \left( \frac{\epsilon_{e,\rm GRB}}{10^{-1}} \right)^{\frac{3p-5}{3}}\ \left( \frac{\epsilon_{B,\rm GRB}}{10^{-2}} \right)^{\frac{3p-2}{12}}\ \\
    &\times\ \left(\frac{\epsilon_{e,\rm SN}}{10^{-1}}\right)^{-\frac{5}{7}}
    \left(\frac{\epsilon_{B,\rm SN}}{10^{-2}}\right)^{-\frac{9}{14}} \left(\frac{V_{\rm sh}}{10^9\,{\rm cm\,s}^{-1}}\right)^{-\frac{19}{17}}.
    \label{eq:fpeak/fp}
\end{split}
\end{equation}
Both ratios depend strongly on the viewing angle, $\theta_{\rm obs}$. In contrast, the ratio of the peak times is independent of the GRB microphysical parameters.

For small values of $\theta_{\rm obs}/\theta_{\rm jet}$, the peak time and peak flux ratios given in Eqs.~(\ref{eq:tm/tp}) and (\ref{eq:fm/fp}) indicate that double-peaked structures are detectable only when a low-luminosity GRB explodes in a dense CSM environment. In this regime, we expect the first peak to be dominated by the GRB component, with the second peak arising from the radio SN emission.
For large $\theta_{\rm obs}/\theta_{\rm jet}$, the peak time and peak flux ratios given in Eqs.~(\ref{eq:tpeak/tp}) and (\ref{eq:fpeak/fp}) show that the conditions for producing double-peaked structures depend strongly on the viewing angle $\theta_{\rm obs}$. In such cases, a second peak from an off-axis radio afterglow may appear up to an order of magnitude later than the first peak from the radio SN. This highlights the importance of long-term radio monitoring of IcBL SNe in order to identify off-axis GRBs.

\section{Diversity of Radio Light Curves} \label{sec:result}

In this section, we systematically study a diversity of radio light curves contributed by SNe and GRB afterglows. 
We adopt $\Gamma_0=100$ and $\theta_{\rm jet}=0.05$ rad for the GRB model and $M_{\rm ej}=1.5M_\odot$ and $E_{\rm ej}=4\times10^{51}$ erg for the SN model.
Since the peak time of the GRB radio afterglow strongly depends on the shock dynamics and geometry due to the relativistic Doppler effect \citep{2017MNRAS.472.4953L,2020ApJ...896..166R}, we focus mainly on the diversity of radio afterglow light curves in a wide range of isotropic equivalent energy $E_{\rm iso}$ and viewing angle $\theta_{\rm obs}$ in the following discussion. 

Based on the shock wave dynamics given in Section \ref{sec:mm}, we calculate the energy distribution of non-thermal electrons by Eq. (\ref{eq:Ngamma}) and synchrotron radiation.
To calculate the observed radio flux $F_{\nu}$, we adopt the source distance as $D_{\rm L}=100$ Mpc and $z=0.024$. 
We use $\epsilon_{\rm e}=10^{-1}$, $\epsilon_{\rm B}=10^{-2}$ for both the GRBs and the SNe models for simplicity. Meanwhile, we adopted $p=2.5$ for the GRBs \citep{1998ApJ...497L..17S} and $p=3.0$ for the SNe model \citep{2006ApJ...641.1029C}. 
The independent choice of the spectral index $p$ is due to the different efficiency of the diffusive shock acceleration. GRB has typically large $\gamma_m\approx600\Gamma_{\rm }$ \citep{1998ApJ...497L..17S} so that non-thermal electrons can be efficiently accelerated. However, SN has small $\gamma_m\sim1$ so that diffusive shock acceleration becomes inefficient, resulting in the softening of the particle spectra compared to GRB and theoretical expectation \citep{2012ApJ...758...81M,2013ApJ...762L..24M}. 

The radio SNe are suppressed by SSA in the early stage of the evolution. As the shock wave expands, the SSA frequency shifts to the lower frequency due to the decrease in the number density and magnetic field strength. The radio flux peaks when the SSA frequency $\nu_{\rm a}$ passes the radio frequency: $\nu_{\rm obs}=\nu_{\rm a}$. 

The GRB radio afterglow is also affected by SSA in the early phase of the evolution, then $\nu_{\rm a}$ decreases as the shock wave expands. 
However, the evolution of radio flux is different from radio SNe in the case of the large synchrotron peak frequency $\nu_{\rm m}>\nu_{\rm a}$. 
The radio flux becomes maximum when the peak frequency $\nu_{\rm m}$ passes the radio frequency: $\nu_{\rm obs}=\nu_{\rm m}$. 
However, if the system is off-axis and violates Eq. (\ref{eq:condition}), the peak flux is achieved when the core of the jet becomes visible. 
The evolution of the radio spectrum depends sensitively on the cooling frequency $\nu_{\rm c}$.
The radio flux reaches a local minimum when the spectrum transitions from the fast-cooling to the slow-cooling regime at $\nu_{\rm c}=\nu_{\rm m}$ \citep[e.g.,][]{2022ApJ...940..189F,2023MNRAS.524.6004W}. 
However, \citet{1998ApJ...497L..17S} shows that the radio flux does not reach the local minimum in the same regime. These inconsistencies may arise from different assumptions about the particle energy distribution. 

\subsection{Dependence of isotropic-equivalent energy} \label{sec:energy}

\begin{figure*}
    \centering
    \includegraphics[width=\linewidth]{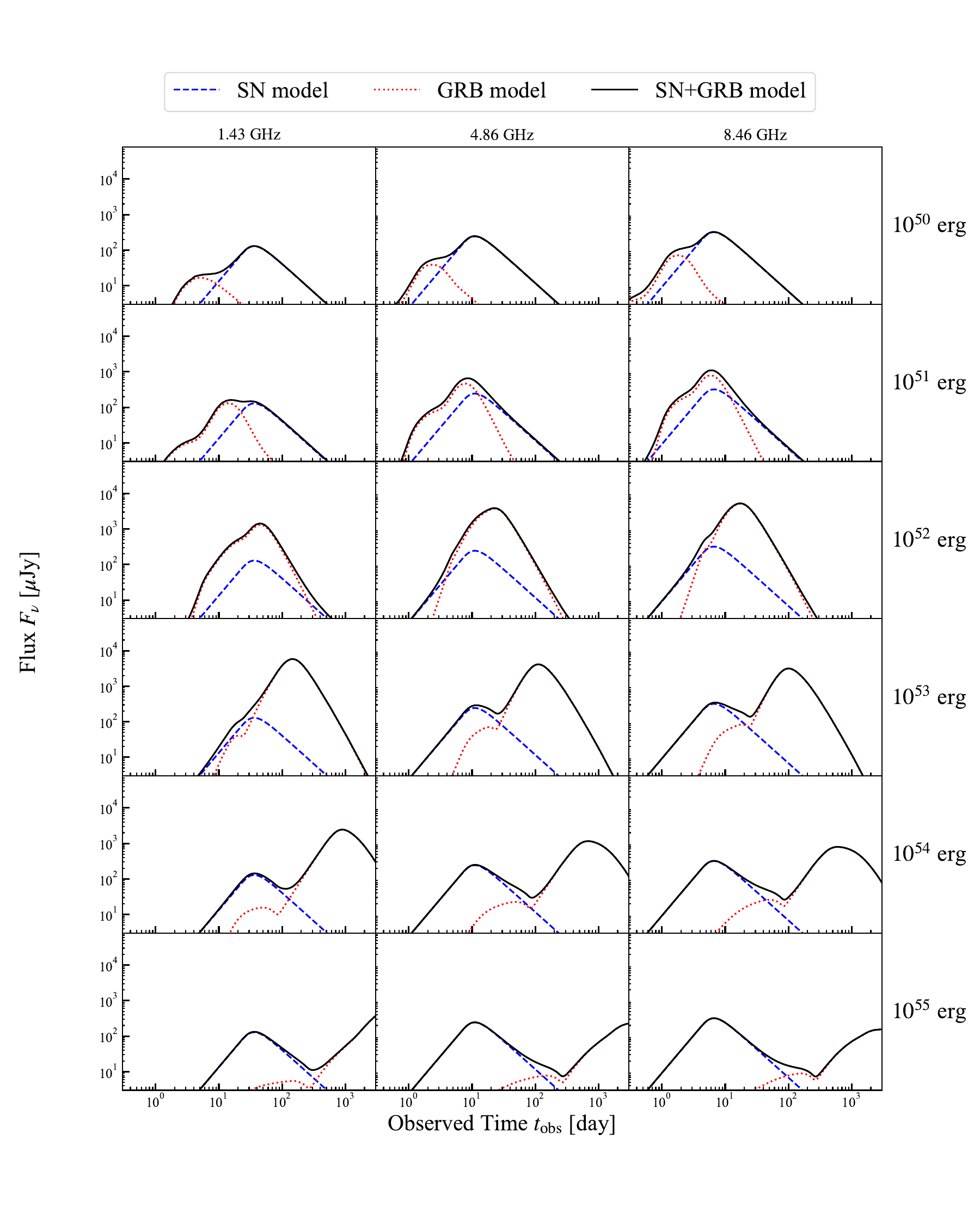}
    \caption{Demonstrative superposition of radio light curves (solid) of an SN (dashed) and a GRB afterglow (dotted) for fixed viewing angle $\theta_{\rm obs}=0.8$ rad. From the top to the bottom, the isotropic-equivalent energy is changed as $E_{\rm iso}=10^{50},\ 10^{51},\ 10^{52},\ 10^{53},\ 10^{54},\ 10^{55}$ erg. }
    \label{fig:demonstration_Eiso}
\end{figure*}

Firstly in Figure \ref{fig:demonstration_Eiso}, we illustrate the radio light curve models in which the components of radio SN and GRB afterglow are superimposed. Here we adopt the CSM density scale of $A=1$, which is derived from the mass-loss rate and wind velocity of Wolf-Rayet stars observed in the Milky Way \citep{2000AandA...360..227N}. 
We calculate light curves of the GRB radio afterglow for a wide range of isotropic-equivalent jet energy $E_{\rm iso}=10^{50}\sim10^{55}$ erg. We fix the viewing angle as $\theta_{\rm obs}=0.8$ rad.

According to Eqs. (\ref{eq:peak}), (\ref{eq:tm}), the peak time of the radio afterglow follows as $\propto E_{\rm iso}^{1/3}$ for the case in which the isotropic-equivalent energy is less than $E_{\rm iso}<10^{50}$ erg.
In the case of $E_{\rm iso}=10^{50}$ erg, the afterglow peak time is $\sim$ 10 times shorter than the SN peak time, resulting in a double-peaked light curve (first peak: GRB $\rightarrow$ second peak: SN). 
If $E_{\rm iso}>10^{51}$ erg, the peak time is determined by $t_{\rm peak}\propto E_{\rm iso}$.
For the $E_{\rm iso}=10^{51}$ and $10^{52}$ erg cases, the peak times of both SN and GRB are comparable, and the resulting light curve becomes a single peak.
For $E_{\rm iso}=10^{53}$ erg, the 1.43 GHz light curve has a single peak, but the 4.86 GHz and the 8.46 GHz are double-peaked. We denote such light curves as "partially double-peaked" in the following.
For the $E_{\rm iso}\geqq10^{54}$ erg case, the afterglow peak time becomes more than 10 times longer than the SN peak time, generating a clear double-peaked light curve (first peak: SN $\rightarrow$ second peak: GRB).  

\subsection{Dependence of viewing angle} \label{sec:theta}

\begin{figure*}
    \centering
    \includegraphics[width=\linewidth]{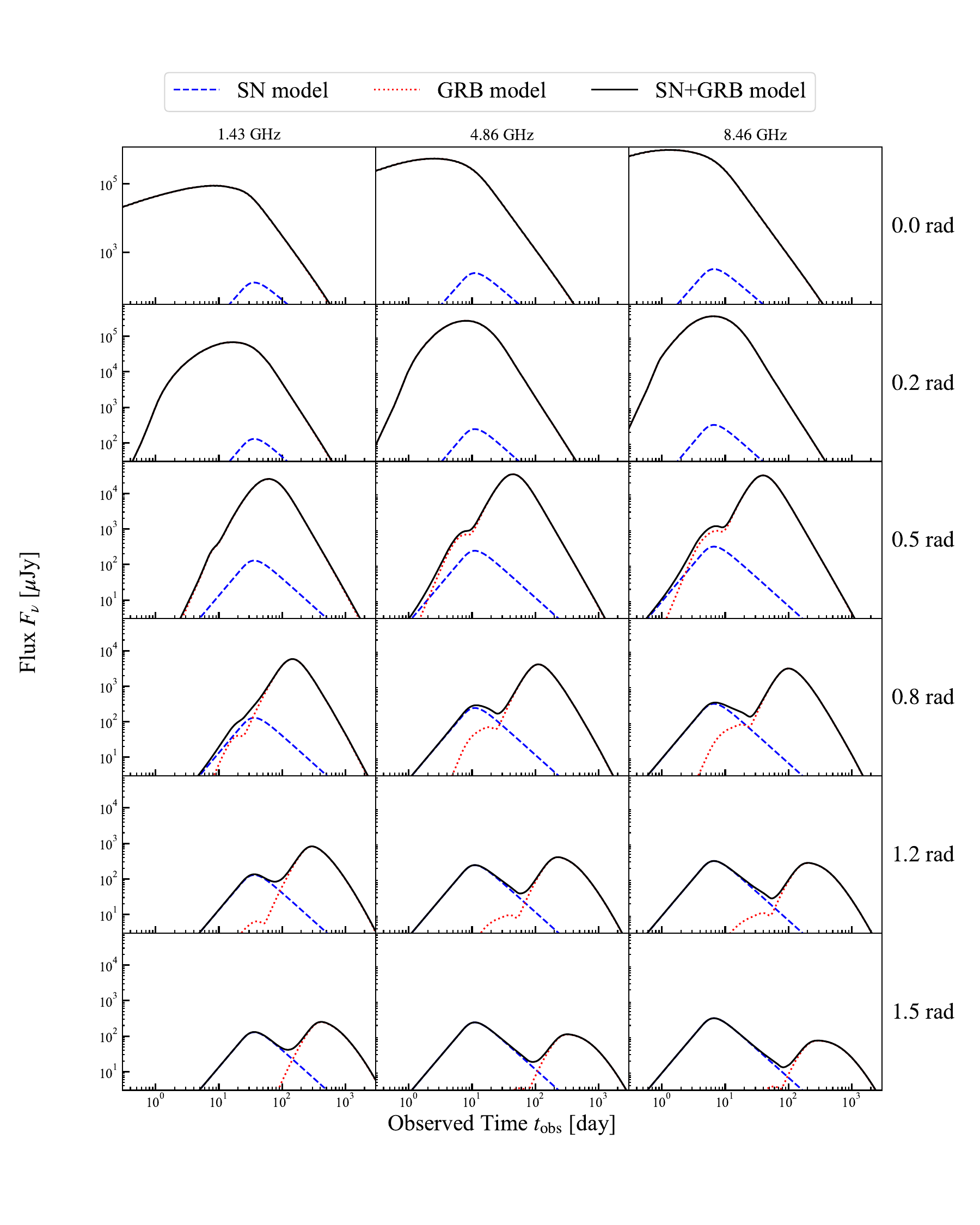}
    \caption{The same as Figure \ref{fig:demonstration_Eiso}, but the isotropic-equivalent energy is fixed to $E_{\rm iso}=10^{53}$ erg. From the top to the bottom, the viewing angle is changed as $\theta_{\rm obs}=$ 0.0, 0.2, 0.5, 0.8, 1.2, 1.5 rad. }
    \label{fig:demonstration_theta}
\end{figure*}

Eq. (\ref{eq:condition}) suggests that the peak time scale strongly depends on the viewing angle $\theta_{\rm obs}$. 
In Figure \ref{fig:demonstration_theta}, we show light curves of the GRB radio afterglow for $A=1$ against a wide range of viewing angle $\theta_{\rm obs}=0.0\sim1.5$ rad for the fixed $E_{\rm iso}=10^{53}$ erg. 
Although the lateral spreading may make the peak time shorter, 
GRB-SN with a larger viewing angle tends to be a double-peaked radio light curve. 

According to Eqs. (\ref{eq:peak}), (\ref{eq:tm}), the peak time is determined by $t_{\rm m}\propto \theta_{\rm obs}^{2/3}$ for $\theta_{\rm obs}<0.3$ rad.
For a small viewing angle ($\theta_{\rm obs}<0.5$ rad, or $\theta_{\rm obs}/\theta_{\rm jet}<10$), the radio SN is overwhelmed by the afterglow component, producing a single-peaked light curve.  
As the viewing angle increases $\theta_{\rm obs}>0.3$ rad, the afterglow peak time becomes longer $t_{\rm peak}\propto\theta_{\rm obs}^4$ and the peak flux decreases $\propto \theta_{\rm obs}^{-3(p-1)}$.
As a result, the radio light curve becomes partially double-peaked, with the radio SN component emerging around $O(10)$ days.  
For larger viewing angles ($\theta_{\rm obs}>1.2$ rad, or $\theta_{\rm obs}/\theta_{\rm jet}>24$),  the afterglow peak time shifts to $O(100)$ days, producing a clear double-peaked light curve (first peak: SN $\rightarrow$ second peak: GRB).  

\subsection{Identification of GRB-SN} \label{sec:obs}


\begin{figure*}
    \centering
    \includegraphics[width=\linewidth]{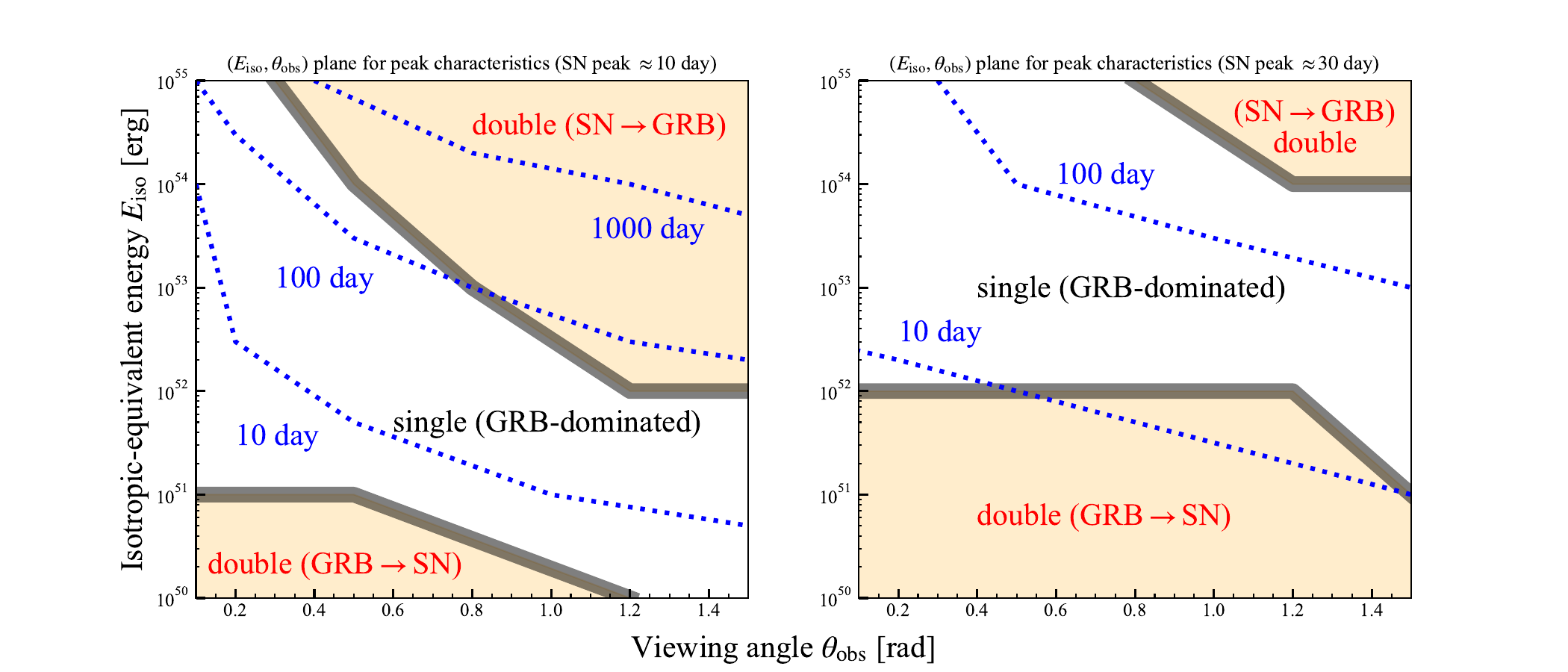}
    \caption{The characteristic $(E_{\rm iso},\theta_{\rm obs})$ plane for $A=1$ (left panel) and $A=10$ (right panel). The blue-dotted lines express the typical peak time of the GRB radio afterglow, meanwhile the typical SN peak is around 10 days. The black shaded regions corresponds to `partial' cases. The fixed parameters are $\Gamma_0=100$, $p_{\rm GRB}=2.5$, $\epsilon_{e,\rm GRB}=10^{-1}$, $\epsilon_{B,\rm GRB}=10^{-2}$, $\theta_{\rm jet}=0.05$ rad. For SN, $p_{\rm SN}=3.0$, $\epsilon_{e,\rm SN}=10^{-1}$ and $\epsilon_{B,\rm SN}=10^{-2}$. The source distance is assumed as $D_{\rm L}=100$ Mpc and $z=0.024$.}
    \label{fig:general}
\end{figure*}

As demonstrated in Figures \ref{fig:demonstration_Eiso}, \ref{fig:demonstration_theta}, the behavior of the light curve has strong dependence on $E_{\rm iso}$ and $\theta_{\rm obs}$. If the condition Eq. (\ref{eq:condition}) is satisfied, the ratio of the peak time and flux given in Eqs. (\ref{eq:tm/tp}), (\ref{eq:fm/fp}) indicate that the double-peaked structure is likely produced for significantly small $E_{\rm iso}$. Since there is no dependence on $\theta_{\rm obs}$, the double peak structure can be seen for both on- and off-axis GRB-SN.
On the other hand, if the condition Eq. (\ref{eq:condition}) is violated, the ratio of the peak time and flux given in Eqs. (\ref{eq:tpeak/tp}), (\ref{eq:fpeak/fp}) suggest that the formation of a double-peaked structure becomes more likely at larger viewing angles, $\theta_{\rm obs}$.

The peak time of the radio SNe and GRB afterglow also depends on the density of the wind bubble.  
Since the radio SN light curve peaks when $\nu_{\rm obs}=\nu_{\rm a}$, the peak time becomes longer for larger $A$. 
According to Eq. (\ref{eq:peak}), the peak time of the afterglow becomes shorter for larger $A$ if the condition Eq. (\ref{eq:condition}) is satisfied. Meanwhile, if the condition is violated, the peak time is independent of $A$.
Thus, double-peaked radio light curves for large $A$ can be produced when the isotropic-equivalent energy $E_{\rm iso}$ and the viewing angle $\theta_{\rm obs}$ are large (first peak: SN $\rightarrow$ second peak: GRB), or both are significantly small (first peak: GRB $\rightarrow$ second peak: SN). 

Combining the above results, we classify the conditions of a single- or double-peaked light curve in the $(E_{\rm iso},\theta_{\rm obs})$ plane. 
Figure \ref{fig:general} shows the characteristic planes for the cases of $A=1$ and $A=10$. 
We classified radio light curves as "double" when it has clear double peak in all 1.43, 4.86, and 8.46 GHz bands and as "single" when it has no double peak in all of them. 
Otherwise, we classified them as "partial". 

We find that the double peak radio light curve whose peaks are at $\mathcal{O}(10)$ days (SNe) and $\mathcal{O}(100)$ days (GRBs) are likely formed when the angle ratio $\theta_{\rm obs}/\theta_{\rm jet}$ is greater than $\sim10$ for the typical GRBs with the isotropic-equivalent energy $E_{\rm iso}=10^{52}\sim10^{53}$ erg for the case of $A=1$, and much larger $E_{\rm iso}$ for the case of $A=10$. If $E_{\rm iso}$ is significantly large (say more than $10^{54}$ erg), the typical peak time of the GRB radio afterglow can be $\mathcal{O}(1000)$ days. 
Thus, the follow-up observation of radio SNe in a few years later might find off-axis GRBs with significantly high $E_{\rm iso}$ comparable to GRB 221009A \citep{2023Sci...380.1390L}. 

On the other hand, we also find that the double peak radio light curve whose peaks are at $\mathcal{O}(1)$ days (GRBs) and $\mathcal{O}(10)$ days (SNe) are likely formed when the isotropic-equivalent energy $E_{\rm iso}$ is less than $10^{52}$ erg, but almost independent of the angle ratio $\theta_{\rm obs}/\theta_{\rm jet}$. 
Thus, rapid follow-up of radio observation for IcBL SNe might find the off-axis low-luminosity GRBs.

\section{Summary and Discussion} \label{sec:discussion}

The discovery of SN~1998bw associated with GRB 980425 has confirmed the plausible association between a Type IcBL SN and a long GRB.
It is naturally expected that both the GRB jet and SN shock propagate in the same CSM produced by their progenitor wind mass-loss. 
Due to the different dynamical and spectral timescales for GRB and SN, the radio light curves will have distinct features, which might be used for exploration for off-axis GRBs. 
Radio observations have been examined to give constraints on the probable associations of off-axis GRBs and Type IcBL SNe, while there have been few attempts to comprehensively predict the morphology of the radio light curve in GRB-SNe. 

In this research, we construct a model for radio light curves of GRB-SNe as a composite of contributions from radio SN and GRB afterglow, paying attention to the expected unified circumstellar environment in a GRB-SN.
We apply our model to SN 2007bg whose radio light curves have double peaks at ${O(100)}$ days and $O(1000)$ days. 
The radio light curves can be explained with reasonable parameters.
The inferred large wind density parameter $A$ may suggest that a progenitor of GRB-associated SNe may experience the violent mass-loss history a few years before the explosion. 
Our result indicates that radio follow-up observations at least a few years after the explosion are required to fully constrain the possible association of off-axis GRBs.

We provide analytic estimates for the peak time and peak flux ratios of radio SNe and off-axis GRB afterglows in order to evaluate the conditions under which a double-peaked structure is produced. The characteristic peak time and peak flux ratios exhibit substantial diversity as a function of the viewing angle ratio $\theta_{\rm obs}/\theta_{\rm jet}$. For small viewing angle ratios, a double peak arises only when the GRB is low-luminosity. In contrast, for large viewing angle ratios, highly inclined off-axis GRBs may become detectable through sustained, multi-year radio monitoring of IcBL SNe.

We also survey the wide ranges of the isotropic-equivalent energy and viewing angle to cover parameter spaces reproducing low-luminosity GRBs and off-axis GRBs. We find the contribution of off-axis GRBs afterglow can produce double-peaked radio light curves of GRB-SNe in specific parameter spaces, the properties of which are described as follows.
\begin{enumerate}
    \item The large isotropic-equivalent energy $E_{\rm iso}>10^{53}$ erg or the large viewing angle ratio $\theta_{\rm obs}/\theta_{\rm jet}>10$ leads to a double-peaked radio light curve, whose first and second peak can originate from SNe and GRBs, respectively.
    \item The small isotropic-equivalent energy $E_{\rm iso}<10^{52}$ erg leads to a double-peaked radio light curve almost independent of the viewing angle $\theta_{\rm obs}$, whose first and second peak can originate from GRBs and SNe, respectively.
\end{enumerate}

It would be useful for us to describe how to tell from which radio-emitting source each peak in the double-peaked light curve is characterized. It would be worth noting that the decay rates of the GRB afterglow are more rapid than that of the radio SN component, as demonstrated in Figure~\ref{fig:demonstration_Eiso} and \ref{fig:demonstration_theta}. This is caused by the rapid deceleration rate of the GRB jet in the Sedov-Taylor phase observed in the late phase. We suppose that the measurement of the decay rate would be a key to discriminating the radio-emitting source between radio SN and GRB afterglow.

In this study, we have neglected the contribution from the counter jet. As the viewing angle $\theta_{\rm obs}$ increases, emission from the counter jet may enhance the observed radio flux, potentially producing a small bump in the radio light curve after the main afterglow peak. In the limiting case of $\theta_{\rm obs} = \pi/2$, the radio afterglow flux from a symmetric two-sided jet would be doubled.
Additional effects, such as jet structure \citep[e.g.,][]{2020ApJ...896..166R,2024JHEAp..41....1O}, may also influence the late-time radio afterglow. A detailed treatment of these effects is beyond the scope of this work and is left for future studies.

Finally, we mention the assumption of microphysics parameters $\epsilon_e$ and $\epsilon_B$ imposed in this study. For the purpose of demonstration in Section \ref{sec:result}, we have just assumed that the values of $\epsilon_e$ and $\epsilon_B$ are equal between radio SN and GRB afterglow each other without introducing any time dependence.
Although the exact values of these microphysics parameters are still unknown, it should be noted that lower values of $\epsilon_e$ are also inferred from previous works \citep{2023MNRAS.518.1522D}. \cite{2024ApJ...970..141A} demonstrated modeling for GRB afterglows with decaying $\epsilon_e$ and $\epsilon_B$ introduced to reproduce typical X-ray afterglow light curves. These effects may be significant in the practical fitting of double-peaked radio light curves of GRB-SNe obtained in the future.
According to our results, the radio light curves of Type IcBL SNe hosting off-axis GRB events can be far more diverse than previously appreciated. In particular, the second peak can emerge on a much longer timescale than the first, as illustrated in Figures~\ref{fig:demonstration_Eiso} and \ref{fig:demonstration_theta}. In contrast, radio follow-up observations of Type IcBL SNe have typically been confined to within about one year after explosion and have rarely extended over several years. Only recently has the importance of late-time radio follow-up, in the context of re-brightening, been recognized \citep{2024MNRAS.534.3853R}.
We suggest that long-term radio monitoring of Type IcBL SNe will be crucial for testing the ubiquity of GRB jets and will provide important clues to the nature of GRB–SN progenitors.
The authors thank the anonymous referee for thoughtful discussions and comments on this work. 
The authors thankfully acknowledge the computer resources provided by the Institute for Cosmic Ray Research (ICRR), the University of Tokyo. 
This work is supported by the joint research program of ICRR, JSPS KAKENHI Grant Numbers JP23KJ0692 (Y.K.) and 21K13964 (R.S.), the National Science and Technology Council, Taiwan under grant No. MOST 110-2112-M-001-068-MY3, and the Academia Sinica, Taiwan under a career development award under grant No. AS-CDA-111-M04 (T.M.).

\bibliography{main}{}

\begin{thebibliography}{}
\expandafter\ifx\csname natexlab\endcsname\relax\def\natexlab#1{#1}\fi
\providecommand{\url}[1]{\href{#1}{#1}}
\providecommand{\dodoi}[1]{doi:~\href{http://doi.org/#1}{\nolinkurl{#1}}}
\providecommand{\doeprint}[1]{\href{http://ascl.net/#1}{\nolinkurl{http://ascl.net/#1}}}
\providecommand{\doarXiv}[1]{\href{https://arxiv.org/abs/#1}{\nolinkurl{https://arxiv.org/abs/#1}}}

\bibitem[{{Aharonian} {et~al.}(2010){Aharonian}, {Kelner}, \& {Prosekin}}]{2010PhRvD..82d3002A}
{Aharonian}, F.~A., {Kelner}, S.~R., \& {Prosekin}, A.~Y. 2010, \prd, 82, 043002, \dodoi{10.1103/PhysRevD.82.043002}

\bibitem[{{Ajello} {et~al.}(2020){Ajello}, {Arimoto}, {Axelsson}, {Baldini}, {Barbiellini}, {Bastieri}, {Bellazzini}, {Berretta}, {Bissaldi}, {Blandford}, {Bonino}, {Bottacini}, {Bregeon}, {Bruel}, {Buehler}, {Burns}, {Buson}, {Cameron}, {Caputo}, {Caraveo}, {Cavazzuti}, {Chen}, {Chiaro}, {Ciprini}, {Cohen-Tanugi}, {Costantin}, {Cutini}, {D'Ammando}, {DeKlotz}, {de la Torre Luque}, {de Palma}, {Desai}, {Di Lalla}, {Di Venere}, {Fana Dirirsa}, {Fegan}, {Franckowiak}, {Fukazawa}, {Funk}, {Fusco}, {Gargano}, {Gasparrini}, {Giglietto}, {Gill}, {Giordano}, {Giroletti}, {Granot}, {Green}, {Grenier}, {Grondin}, {Guiriec}, {Hays}, {Horan}, {J{\'o}hannesson}, {Kocevski}, {Kovac'evic'}, {Kuss}, {Larsson}, {Latronico}, {Lemoine-Goumard}, {Li}, {Liodakis}, {Longo}, {Loparco}, {Lovellette}, {Lubrano}, {Maldera}, {Malyshev}, {Manfreda}, {Mart{\'\i}-Devesa}, {Mazziotta}, {McEnery}, {Mereu}, {Meyer}, {Michelson}, {Mitthumsiri}, {Mizuno}, {Monzani}, {Moretti}, {Morselli}, {Moskalenko}, {Negro}, {Nuss}, {Omodei}, {Orienti},
  {Orlando}, {Palatiello}, {Paliya}, {Paneque}, {Pei}, {Persic}, {Pesce-Rollins}, {Petrosian}, {Piron}, {Poon}, {Porter}, {Principe}, {Racusin}, {Rain{\`o}}, {Rando}, {Rani}, {Razzano}, {Razzaque}, {Reimer}, {Reimer}, {Ryde}, {Saz Parkinson}, {Serini}, {Sgr{\`o}}, {Siskind}, {Spandre}, {Spinelli}, {Tajima}, {Takagi}, {Takahashi}, {Tak}, {Thayer}, {Thompson}, {Torres}, {Troja}, {Valverde}, {Van Klaveren}, {Wood}, {Yassine}, {Zaharijas}, {Mailyan}, {Bhat}, {Briggs}, {Cleveland}, {Giles}, {Goldstein}, {Hui}, {Malacaria}, {Preece}, {Roberts}, {Veres}, {Wilson-Hodge}, {Kienlin}, {Cenko}, {O'Brien}, {Beardmore}, {Lien}, {Osborne}, {Tohuvavohu}, {D'Elia}, {D'A{\`\i}}, {Perri}, {Gropp}, {Klingler}, {Capalbi}, {Tagliaferri}, {Stamatikos}, \& {De Pasquale}}]{2020ApJ...890....9A}
{Ajello}, M., {Arimoto}, M., {Axelsson}, M., {et~al.} 2020, \apj, 890, 9, \dodoi{10.3847/1538-4357/ab5b05}

\bibitem[{{Anderson} {et~al.}(2014){Anderson}, {van der Horst}, {Staley}, {Fender}, {Wijers}, {Scaife}, {Rumsey}, {Titterington}, {Rowlinson}, \& {Saunders}}]{2014MNRAS.440.2059A}
{Anderson}, G.~E., {van der Horst}, A.~J., {Staley}, T.~D., {et~al.} 2014, \mnras, 440, 2059, \dodoi{10.1093/mnras/stu478}

\bibitem[{{Anderson} {et~al.}(2023){Anderson}, {Russell}, {Fausey}, {van der Horst}, {Hancock}, {Bahramian}, {Bell}, {Miller-Jones}, {Rowell}, {Sammons}, {Wijers}, {Galvin}, {Goodwin}, {Konno}, {Rowlinson}, {Ryder}, {Sch{\"u}ssler}, {Wagner}, \& {Zhu}}]{2023MNRAS.523.4992A}
{Anderson}, G.~E., {Russell}, T.~D., {Fausey}, H.~M., {et~al.} 2023, \mnras, 523, 4992, \dodoi{10.1093/mnras/stad1635}

\bibitem[{{Asano}(2024)}]{2024ApJ...970..141A}
{Asano}, K. 2024, \apj, 970, 141, \dodoi{10.3847/1538-4357/ad6148}

\bibitem[{{Asano} {et~al.}(2014){Asano}, {Takahara}, {Kusunose}, {Toma}, \& {Kakuwa}}]{2014ApJ...780...64A}
{Asano}, K., {Takahara}, F., {Kusunose}, M., {Toma}, K., \& {Kakuwa}, J. 2014, \apj, 780, 64, \dodoi{10.1088/0004-637X/780/1/64}

\bibitem[{{Asano} \& {Terasawa}(2009)}]{2009ApJ...705.1714A}
{Asano}, K., \& {Terasawa}, T. 2009, \apj, 705, 1714, \dodoi{10.1088/0004-637X/705/2/1714}

\bibitem[{{Ashall} {et~al.}(2018){Ashall}, {Mazzali}, {Stritzinger}, {Hoeflich}, {Burns}, {Gall}, {Hsiao}, {Phillips}, {Morrell}, \& {Foley}}]{2018MNRAS.477..153A}
{Ashall}, C., {Mazzali}, P.~A., {Stritzinger}, M.~D., {et~al.} 2018, \mnras, 477, 153, \dodoi{10.1093/mnras/sty632}

\bibitem[{{Bell}(1978)}]{1978MNRAS.182..147B}
{Bell}, A.~R. 1978, \mnras, 182, 147, \dodoi{10.1093/mnras/182.2.147}

\bibitem[{{Beniamini} {et~al.}(2020){Beniamini}, {Granot}, \& {Gill}}]{2020MNRAS.493.3521B}
{Beniamini}, P., {Granot}, J., \& {Gill}, R. 2020, \mnras, 493, 3521, \dodoi{10.1093/mnras/staa538}

\bibitem[{{Berger} {et~al.}(2011){Berger}, {Chornock}, {Holmes}, {Foley}, {Cucchiara}, {Wolf}, {Podsiadlowski}, {Fox}, \& {Roth}}]{2011ApJ...743..204B}
{Berger}, E., {Chornock}, R., {Holmes}, T.~R., {et~al.} 2011, \apj, 743, 204, \dodoi{10.1088/0004-637X/743/2/204}

\bibitem[{{Bersier} {et~al.}(2006){Bersier}, {Fruchter}, {Strolger}, {Gorosabel}, {Levan}, {Burud}, {Rhoads}, {Becker}, {Cassan}, {Chornock}, {Covino}, {de Jong}, {Dominis}, {Filippenko}, {Hjorth}, {Holmberg}, {Malesani}, {Mobasher}, {Olsen}, {Stefanon}, {Castro Cer{\'o}n}, {Fynbo}, {Holland}, {Kouveliotou}, {Pedersen}, {Tanvir}, \& {Woosley}}]{2006ApJ...643..284B}
{Bersier}, D., {Fruchter}, A.~S., {Strolger}, L.~G., {et~al.} 2006, \apj, 643, 284, \dodoi{10.1086/502640}

\bibitem[{{Bhirombhakdi} {et~al.}(2024){Bhirombhakdi}, {Fruchter}, {Levan}, {Pian}, {Mazzali}, {Izzo}, {Kangas}, {Benetti}, {Medler}, \& {Tanvir}}]{2024ApJ...977..256B}
{Bhirombhakdi}, K., {Fruchter}, A.~S., {Levan}, A.~J., {et~al.} 2024, \apj, 977, 256, \dodoi{10.3847/1538-4357/ad8dd8}

\bibitem[{{Bietenholz} {et~al.}(2014){Bietenholz}, {De Colle}, {Granot}, {Bartel}, \& {Soderberg}}]{2014MNRAS.440..821B}
{Bietenholz}, M.~F., {De Colle}, F., {Granot}, J., {Bartel}, N., \& {Soderberg}, A.~M. 2014, \mnras, 440, 821, \dodoi{10.1093/mnras/stu246}

\bibitem[{{Bj{\"o}rnsson}(2022)}]{2022ApJ...936...98B}
{Bj{\"o}rnsson}, C.~I. 2022, \apj, 936, 98, \dodoi{10.3847/1538-4357/ac87aa}

\bibitem[{{Blandford} \& {McKee}(1976)}]{1976PhFl...19.1130B}
{Blandford}, R.~D., \& {McKee}, C.~F. 1976, Physics of Fluids, 19, 1130, \dodoi{10.1063/1.861619}

\bibitem[{{Blandford} \& {Ostriker}(1978)}]{1978ApJ...221L..29B}
{Blandford}, R.~D., \& {Ostriker}, J.~P. 1978, \apjl, 221, L29, \dodoi{10.1086/182658}

\bibitem[{{Bloom} {et~al.}(2002){Bloom}, {Kulkarni}, {Price}, {Reichart}, {Galama}, {Schmidt}, {Frail}, {Berger}, {McCarthy}, {Chevalier}, {Wheeler}, {Halpern}, {Fox}, {Djorgovski}, {Harrison}, {Sari}, {Axelrod}, {Kimble}, {Holtzman}, {Hurley}, {Frontera}, {Piro}, \& {Costa}}]{2002ApJ...572L..45B}
{Bloom}, J.~S., {Kulkarni}, S.~R., {Price}, P.~A., {et~al.} 2002, \apjl, 572, L45, \dodoi{10.1086/341551}

\bibitem[{{Bright} {et~al.}(2019){Bright}, {Horesh}, {van der Horst}, {Fender}, {Anderson}, {Motta}, {Cenko}, {Green}, {Perrott}, \& {Titterington}}]{2019MNRAS.486.2721B}
{Bright}, J.~S., {Horesh}, A., {van der Horst}, A.~J., {et~al.} 2019, \mnras, 486, 2721, \dodoi{10.1093/mnras/stz1004}

\bibitem[{{Campana} {et~al.}(2006){Campana}, {Mangano}, {Blustin}, {Brown}, {Burrows}, {Chincarini}, {Cummings}, {Cusumano}, {Della Valle}, {Malesani}, {M{\'e}sz{\'a}ros}, {Nousek}, {Page}, {Sakamoto}, {Waxman}, {Zhang}, {Dai}, {Gehrels}, {Immler}, {Marshall}, {Mason}, {Moretti}, {O'Brien}, {Osborne}, {Page}, {Romano}, {Roming}, {Tagliaferri}, {Cominsky}, {Giommi}, {Godet}, {Kennea}, {Krimm}, {Angelini}, {Barthelmy}, {Boyd}, {Palmer}, {Wells}, \& {White}}]{2006Natur.442.1008C}
{Campana}, S., {Mangano}, V., {Blustin}, A.~J., {et~al.} 2006, \nat, 442, 1008, \dodoi{10.1038/nature04892}

\bibitem[{{Cano} {et~al.}(2017){Cano}, {Wang}, {Dai}, \& {Wu}}]{2017AdAst2017E...5C}
{Cano}, Z., {Wang}, S.-Q., {Dai}, Z.-G., \& {Wu}, X.-F. 2017, Advances in Astronomy, 2017, 8929054, \dodoi{10.1155/2017/8929054}

\bibitem[{{Cano} {et~al.}(2014){Cano}, {de Ugarte Postigo}, {Pozanenko}, {Butler}, {Th{\"o}ne}, {Guidorzi}, {Kr{\"u}hler}, {Gorosabel}, {Jakobsson}, {Leloudas}, {Malesani}, {Hjorth}, {Melandri}, {Mundell}, {Wiersema}, {D'Avanzo}, {Schulze}, {Gomboc}, {Johansson}, {Zheng}, {Kann}, {Knust}, {Varela}, {Akerlof}, {Bloom}, {Burkhonov}, {Cooke}, {de Diego}, {Dhungana}, {Farina}, {Ferrante}, {Flewelling}, {Fox}, {Fynbo}, {Gehrels}, {Georgiev}, {Gonz{\'a}lez}, {Greiner}, {G{\"u}ver}, {Hartoog}, {Hatch}, {Jelinek}, {Kehoe}, {Klose}, {Klunko}, {Kopa{\v{c}}}, {Kutyrev}, {Krugly}, {Lee}, {Levan}, {Linkov}, {Matkin}, {Minikulov}, {Molotov}, {Prochaska}, {Richer}, {Rom{\'a}n-Z{\'u}{\~n}iga}, {Rumyantsev}, {S{\'a}nchez-Ram{\'\i}rez}, {Steele}, {Tanvir}, {Volnova}, {Watson}, {Xu}, \& {Yuan}}]{2014AandA...568A..19C}
{Cano}, Z., {de Ugarte Postigo}, A., {Pozanenko}, A., {et~al.} 2014, \aap, 568, A19, \dodoi{10.1051/0004-6361/201423920}

\bibitem[{{Cano} {et~al.}(2015){Cano}, {de Ugarte Postigo}, {Perley}, {Kr{\"u}hler}, {Margutti}, {Friis}, {Malesani}, {Jakobsson}, {Fynbo}, {Gorosabel}, {Hjorth}, {S{\'a}nchez-Ram{\'\i}rez}, {Schulze}, {Tanvir}, {Th{\"o}ne}, \& {Xu}}]{2015MNRAS.452.1535C}
{Cano}, Z., {de Ugarte Postigo}, A., {Perley}, D., {et~al.} 2015, \mnras, 452, 1535, \dodoi{10.1093/mnras/stv1327}

\bibitem[{{Chen} {et~al.}(2021){Chen}, {Urata}, \& {Huang}}]{2021ApJ...915...46C}
{Chen}, J.-C., {Urata}, Y., \& {Huang}, K. 2021, \apj, 915, 46, \dodoi{10.3847/1538-4357/ac00b4}

\bibitem[{{Chevalier}(1982{\natexlab{a}})}]{1982ApJ...259..302C}
{Chevalier}, R.~A. 1982{\natexlab{a}}, \apj, 259, 302, \dodoi{10.1086/160167}

\bibitem[{{Chevalier}(1982{\natexlab{b}})}]{1982ApJ...258..790C}
---. 1982{\natexlab{b}}, \apj, 258, 790, \dodoi{10.1086/160126}

\bibitem[{{Chevalier}(1998)}]{1998ApJ...499..810C}
---. 1998, \apj, 499, 810, \dodoi{10.1086/305676}

\bibitem[{{Chevalier} \& {Fransson}(1994)}]{1994ApJ...420..268C}
{Chevalier}, R.~A., \& {Fransson}, C. 1994, \apj, 420, 268, \dodoi{10.1086/173557}

\bibitem[{{Chevalier} \& {Fransson}(2017)}]{2017hsn..book..875C}
---. 2017, in Handbook of Supernovae, ed. A.~W. {Alsabti} \& P.~{Murdin}, 875, \dodoi{10.1007/978-3-319-21846-5_34}

\bibitem[{{Chevalier} {et~al.}(2006){Chevalier}, {Fransson}, \& {Nymark}}]{2006ApJ...641.1029C}
{Chevalier}, R.~A., {Fransson}, C., \& {Nymark}, T.~K. 2006, \apj, 641, 1029, \dodoi{10.1086/500528}

\bibitem[{{Cobb} {et~al.}(2010){Cobb}, {Bloom}, {Perley}, {Morgan}, {Cenko}, \& {Filippenko}}]{2010ApJ...718L.150C}
{Cobb}, B.~E., {Bloom}, J.~S., {Perley}, D.~A., {et~al.} 2010, \apjl, 718, L150, \dodoi{10.1088/2041-8205/718/2/L150}

\bibitem[{{Corsi} {et~al.}(2014){Corsi}, {Ofek}, {Gal-Yam}, {Frail}, {Kulkarni}, {Fox}, {Kasliwal}, {Sullivan}, {Horesh}, {Carpenter}, {Maguire}, {Arcavi}, {Cenko}, {Cao}, {Mooley}, {Pan}, {Sesar}, {Sternberg}, {Xu}, {Bersier}, {James}, {Bloom}, \& {Nugent}}]{2014ApJ...782...42C}
{Corsi}, A., {Ofek}, E.~O., {Gal-Yam}, A., {et~al.} 2014, \apj, 782, 42, \dodoi{10.1088/0004-637X/782/1/42}

\bibitem[{{Corsi} {et~al.}(2016){Corsi}, {Gal-Yam}, {Kulkarni}, {Frail}, {Mazzali}, {Cenko}, {Kasliwal}, {Cao}, {Horesh}, {Palliyaguru}, {Perley}, {Laher}, {Taddia}, {Leloudas}, {Maguire}, {Nugent}, {Sollerman}, \& {Sullivan}}]{2016ApJ...830...42C}
{Corsi}, A., {Gal-Yam}, A., {Kulkarni}, S.~R., {et~al.} 2016, \apj, 830, 42, \dodoi{10.3847/0004-637X/830/1/42}

\bibitem[{{Corsi} {et~al.}(2023){Corsi}, {Ho}, {Cenko}, {Kulkarni}, {Anand}, {Yang}, {Sollerman}, {Srinivasaragavan}, {Omand}, {Balasubramanian}, {Frail}, {Fremling}, {Perley}, {Yao}, {Dahiwale}, {De}, {Dugas}, {Hankins}, {Jencson}, {Kasliwal}, {Tzanidakis}, {Bellm}, {Laher}, {Masci}, {Purdum}, \& {Regnault}}]{2023ApJ...953..179C}
{Corsi}, A., {Ho}, A. Y.~Q., {Cenko}, S.~B., {et~al.} 2023, \apj, 953, 179, \dodoi{10.3847/1538-4357/acd3f2}

\bibitem[{{De Colle} {et~al.}(2018){De Colle}, {Kumar}, \& {Aguilera-Dena}}]{2018ApJ...863...32D}
{De Colle}, F., {Kumar}, P., \& {Aguilera-Dena}, D.~R. 2018, \apj, 863, 32, \dodoi{10.3847/1538-4357/aad04d}

\bibitem[{{De Colle} {et~al.}(2022){De Colle}, {Kumar}, \& {Hoeflich}}]{2022MNRAS.512.3627D}
{De Colle}, F., {Kumar}, P., \& {Hoeflich}, P. 2022, \mnras, 512, 3627, \dodoi{10.1093/mnras/stac742}

\bibitem[{{D'Elia} {et~al.}(2015){D'Elia}, {Pian}, {Melandri}, {D'Avanzo}, {Della Valle}, {Mazzali}, {Piranomonte}, {Tagliaferri}, {Antonelli}, {Bufano}, {Covino}, {Fugazza}, {Malesani}, {M{\o}ller}, \& {Palazzi}}]{2015AandA...577A.116D}
{D'Elia}, V., {Pian}, E., {Melandri}, A., {et~al.} 2015, \aap, 577, A116, \dodoi{10.1051/0004-6361/201425381}

\bibitem[{{Della Valle} {et~al.}(2003){Della Valle}, {Malesani}, {Benetti}, {Testa}, {Hamuy}, {Antonelli}, {Chincarini}, {Cocozza}, {Covino}, {D'Avanzo}, {Fugazza}, {Ghisellini}, {Gilmozzi}, {Lazzati}, {Mason}, {Mazzali}, \& {Stella}}]{2003AandA...406L..33D}
{Della Valle}, M., {Malesani}, D., {Benetti}, S., {et~al.} 2003, \aap, 406, L33, \dodoi{10.1051/0004-6361:20030855}

\bibitem[{{Duffell} \& {Laskar}(2018)}]{2018ApJ...865...94D}
{Duffell}, P.~C., \& {Laskar}, T. 2018, \apj, 865, 94, \dodoi{10.3847/1538-4357/aadb9c}

\bibitem[{{Duncan} {et~al.}(2023){Duncan}, {van der Horst}, \& {Beniamini}}]{2023MNRAS.518.1522D}
{Duncan}, R.~A., {van der Horst}, A.~J., \& {Beniamini}, P. 2023, \mnras, 518, 1522, \dodoi{10.1093/mnras/stac3172}

\bibitem[{{Fan} \& {Piran}(2006)}]{2006MNRAS.369..197F}
{Fan}, Y., \& {Piran}, T. 2006, \mnras, 369, 197, \dodoi{10.1111/j.1365-2966.2006.10280.x}

\bibitem[{{Filgas} {et~al.}(2011){Filgas}, {Greiner}, {Schady}, {Kr{\"u}hler}, {Updike}, {Klose}, {Nardini}, {Kann}, {Rossi}, {Sudilovsky}, {Afonso}, {Clemens}, {Elliott}, {Nicuesa Guelbenzu}, {Olivares E.}, \& {Rau}}]{2011AandA...535A..57F}
{Filgas}, R., {Greiner}, J., {Schady}, P., {et~al.} 2011, \aap, 535, A57, \dodoi{10.1051/0004-6361/201117695}

\bibitem[{{Finneran} {et~al.}(2024){Finneran}, {Cotter}, \& {Martin-Carrillo}}]{2024arXiv241108866F}
{Finneran}, G., {Cotter}, L., \& {Martin-Carrillo}, A. 2024, arXiv e-prints, arXiv:2411.08866, \dodoi{10.48550/arXiv.2411.08866}

\bibitem[{{Finneran} {et~al.}(2025){Finneran}, {Cotter}, \& {Martin-Carrillo}}]{2025A&C....5200954F}
---. 2025, Astronomy and Computing, 52, 100954, \dodoi{10.1016/j.ascom.2025.100954}

\bibitem[{{Fouka} \& {Ouichaoui}(2013)}]{2013RAA....13..680F}
{Fouka}, M., \& {Ouichaoui}, S. 2013, Research in Astronomy and Astrophysics, 13, 680, \dodoi{10.1088/1674-4527/13/6/007}

\bibitem[{{Fox} {et~al.}(2003){Fox}, {Price}, {Soderberg}, {Berger}, {Kulkarni}, {Sari}, {Frail}, {Harrison}, {Yost}, {Matthews}, {Peterson}, {Tanaka}, {Christiansen}, \& {Moriarty-Schieven}}]{2003ApJ...586L...5F}
{Fox}, D.~W., {Price}, P.~A., {Soderberg}, A.~M., {et~al.} 2003, \apjl, 586, L5, \dodoi{10.1086/374683}

\bibitem[{{Fraija} {et~al.}(2022){Fraija}, {Galvan-Gamez}, {Betancourt Kamenetskaia}, {Dainotti}, {Dichiara}, {Veres}, {Becerra}, \& {do E.~S. Pedreira}}]{2022ApJ...940..189F}
{Fraija}, N., {Galvan-Gamez}, A., {Betancourt Kamenetskaia}, B., {et~al.} 2022, \apj, 940, 189, \dodoi{10.3847/1538-4357/ac68e1}

\bibitem[{{Fransson} \& {Bj{\"o}rnsson}(1998)}]{1998ApJ...509..861F}
{Fransson}, C., \& {Bj{\"o}rnsson}, C.-I. 1998, \apj, 509, 861, \dodoi{10.1086/306531}

\bibitem[{{Galama} {et~al.}(1999){Galama}, {Vreeswijk}, {van Paradijs}, {Kouveliotou}, {Augusteijn}, {Patat}, {Heise}, {in 't Zand}, {Groot}, {Wijers}, {Pian}, {Palazzi}, {Frontera}, \& {Masetti}}]{1999AandAS..138..465G}
{Galama}, T.~J., {Vreeswijk}, P.~M., {van Paradijs}, J., {et~al.} 1999, \aaps, 138, 465, \dodoi{10.1051/aas:1999311}

\bibitem[{{Gao} {et~al.}(2013){Gao}, {Lei}, {Zou}, {Wu}, \& {Zhang}}]{2013NewAR..57..141G}
{Gao}, H., {Lei}, W.-H., {Zou}, Y.-C., {Wu}, X.-F., \& {Zhang}, B. 2013, \nar, 57, 141, \dodoi{10.1016/j.newar.2013.10.001}

\bibitem[{{Giannios} {et~al.}(2008){Giannios}, {Mimica}, \& {Aloy}}]{2008AandA...478..747G}
{Giannios}, D., {Mimica}, P., \& {Aloy}, M.~A. 2008, \aap, 478, 747, \dodoi{10.1051/0004-6361:20078931}

\bibitem[{{Granot}(2005)}]{2005ApJ...631.1022G}
{Granot}, J. 2005, \apj, 631, 1022, \dodoi{10.1086/432676}

\bibitem[{{Granot} {et~al.}(2003){Granot}, {Nakar}, \& {Piran}}]{2003Natur.426..138G}
{Granot}, J., {Nakar}, E., \& {Piran}, T. 2003, \nat, 426, 138, \dodoi{10.1038/426138a}

\bibitem[{{Granot} {et~al.}(2002){Granot}, {Panaitescu}, {Kumar}, \& {Woosley}}]{2002ApJ...570L..61G}
{Granot}, J., {Panaitescu}, A., {Kumar}, P., \& {Woosley}, S.~E. 2002, \apjl, 570, L61, \dodoi{10.1086/340991}

\bibitem[{{Greiner} {et~al.}(2015){Greiner}, {Mazzali}, {Kann}, {Kr{\"u}hler}, {Pian}, {Prentice}, {Olivares E.}, {Rossi}, {Klose}, {Taubenberger}, {Knust}, {Afonso}, {Ashall}, {Bolmer}, {Delvaux}, {Diehl}, {Elliott}, {Filgas}, {Fynbo}, {Graham}, {Guelbenzu}, {Kobayashi}, {Leloudas}, {Savaglio}, {Schady}, {Schmidl}, {Schweyer}, {Sudilovsky}, {Tanga}, {Updike}, {van Eerten}, \& {Varela}}]{2015Natur.523..189G}
{Greiner}, J., {Mazzali}, P.~A., {Kann}, D.~A., {et~al.} 2015, \nat, 523, 189, \dodoi{10.1038/nature14579}

\bibitem[{{H.~E.~S.~S. Collaboration} {et~al.}(2021){H.~E.~S.~S. Collaboration}, {Abdalla}, {Aharonian}, {Ait Benkhali}, {Ang{\"u}ner}, {Arcaro}, {Armand}, {Armstrong}, {Ashkar}, {Backes}, {Baghmanyan}, {Barbosa Martins}, {Barnacka}, {Barnard}, {Becherini}, {Berge}, {Bernl{\"o}hr}, {Bi}, {Bissaldi}, {B{\"o}ttcher}, {Boisson}, {Bolmont}, {de Bony de Lavergne}, {Breuhaus}, {Brun}, {Brun}, {Bryan}, {B{\"u}chele}, {Bulik}, {Bylund}, {Caroff}, {Carosi}, {Casanova}, {Chand}, {Chandra}, {Chen}, {Cotter}, {Cury{\l}o}, {Damascene Mbarubucyeye}, {Davids}, {Davies}, {Deil}, {Devin}, {Dirson}, {Djannati-Ata{\"\i}}, {Dmytriiev}, {Donath}, {Doroshenko}, {Dreyer}, {Duffy}, {Dyks}, {Egberts}, {Eichhorn}, {Einecke}, {Emery}, {Ernenwein}, {Feijen}, {Fegan}, {Fiasson}, {Fichet de Clairfontaine}, {Fontaine}, {Funk}, {F{\"u}{\ss}ling}, {Gabici}, {Gallant}, {Giavitto}, {Giunti}, {Glawion}, {Glicenstein}, {Grondin}, {Hahn}, {Haupt}, {Hermann}, {Hinton}, {Hofmann}, {Hoischen}, {Holch}, {Holler}, {H{\"o}rbe}, {Horns}, {Huber},
  {Jamrozy}, {Jankowsky}, {Jankowsky}, {Jardin-Blicq}, {Joshi}, {Jung-Richardt}, {Kasai}, {Kastendieck}, {Katarzy{\'n}ski}, {Katz}, {Khangulyan}, {Kh{\'e}lifi}, {Klepser}, {Klu{\'z}niak}, {Komin}, {Konno}, {Kosack}, {Kostunin}, {Kreter}, {Lamanna}, {Lemi{\`e}re}, {Lemoine-Goumard}, {Lenain}, {Leuschner}, {Levy}, {Lohse}, {Lypova}, {Mackey}, {Majumdar}, {Malyshev}, {Malyshev}, {Marandon}, {Marchegiani}, {Marcowith}, {Mares}, {Mart{\'\i}-Devesa}, {Marx}, {Maurin}, {Meintjes}, {Meyer}, {Mitchell}, {Moderski}, {Mohrmann}, {Montanari}, {Moore}, {Morris}, {Moulin}, {Muller}, {Murach}, {Nakashima}, {Nayerhoda}, {de Naurois}, {Ndiyavala}, {Niemiec}, {Oakes}, {O'Brien}, {Odaka}, {Ohm}, {Olivera-Nieto}, {de Ona Wilhelmi}, {Ostrowski}, {Panny}, {Panter}, {Parsons}, {Peron}, {Peyaud}, {Piel}, {Pita}, {Poireau}, {Priyana Noel}, {Prokhorov}, {Prokoph}, {P{\"u}hlhofer}, {Punch}, {Quirrenbach}, {Raab}, {Rauth}, {Reichherzer}, {Reimer}, {Reimer}, {Remy}, {Renaud}, {Rieger}, {Rinchiuso}, {Romoli}, {Rowell}, {Rudak},
  {Ruiz-Velasco}, {Sahakian}, {Sailer}, {Salzmann}, {Sanchez}, {Santangelo}, {Sasaki}, {Scalici}, {Sch{\"a}fer}, {Sch{\"u}ssler}, {Schutte}, {Schwanke}, {Seglar-Arroyo}, {Senniappan}, {Seyffert}, {Shafi}, {Shapopi}, {Shiningayamwe}, {Simoni}, {Sinha}, {Sol}, {Specovius}, {Spencer}, {Spir-Jacob}, {Stawarz}, {Sun}, {Steenkamp}, {Stegmann}, {Steinmassl}, {Steppa}, {Takahashi}, \& {Tam}}]{2021Sci...372.1081H}
{H.~E.~S.~S. Collaboration}, {Abdalla}, H., {Aharonian}, F., {et~al.} 2021, Science, 372, 1081, \dodoi{10.1126/science.abe8560}

\bibitem[{{Hussenot-Desenonges} {et~al.}(2024){Hussenot-Desenonges}, {Wouters}, {Guessoum}, {Abdi}, {Abulwfa}, {Adami}, {Ag{\"u}{\'\i} Fern{\'a}ndez}, {Ahumada}, {Aivazyan}, {Akl}, {Anand}, {Andrade}, {Antier}, {Ata}, {D'Avanzo}, {Azzam}, {Baransky}, {Basa}, {Blazek}, {Bendjoya}, {Beradze}, {Boumis}, {Bremer}, {Brivio}, {Buat}, {Bulla}, {Burkhonov}, {Burns}, {Cenko}, {Coughlin}, {Corradi}, {Daigne}, {Dietrich}, {Dornic}, {Ducoin}, {Duverne}, {Elhosseiny}, {Elnagahy}, {El-Sadek}, {Ferro}, {Le Floc'h}, {Freeberg}, {Fynbo}, {G{\"o}tz}, {Gurbanov}, {Hamed}, {Hasanov}, {Healy}, {Heintz}, {Hello}, {Inasaridze}, {Iskandar}, {Ismailov}, {Izzo}, {Jhawar}, {Jegou du Laz}, {Kamel}, {Karpov}, {Klotz}, {Koulouridis}, {Kuin}, {Kochiashvili}, {Leonini}, {Lu}, {Malesani}, {Ma{\v{s}}ek}, {Mao}, {Melandri}, {Mihov}, {Natsvlishvili}, {Navarete}, {Nedora}, {Nicolas}, {Odeh}, {Palmerio}, {Pang}, {De Pasquale}, {Peng}, {Pormente}, {Peloton}, {Pradier}, {Pyshna}, {Rajabov}, {Rakotondrainibe}, {Rivet}, {Rousselot}, {Saccardi},
  {Sasaki}, {Schneider}, {Serrau}, {Shokry}, {Slavcheva-Mihova}, {Simon}, {Sokoliuk}, {Srinivasaragavan}, {Strausbaugh}, {Takey}, {Tanvir}, {Th{\"o}ne}, {Tillayev}, {Tosta e Melo}, {Turpin}, {de Ugarte Postigo}, {Vasylenko}, {Vergani}, {Vidadi}, {Xu}, {Wang}, {Wang}, {Winters}, {Zhang}, \& {Zhu}}]{2024MNRAS.530....1H}
{Hussenot-Desenonges}, T., {Wouters}, T., {Guessoum}, N., {et~al.} 2024, \mnras, 530, 1, \dodoi{10.1093/mnras/stae503}

\bibitem[{{Izzo} {et~al.}(2020){Izzo}, {Auchettl}, {Hjorth}, {De Colle}, {Gall}, {Angus}, {Raimundo}, \& {Ramirez-Ruiz}}]{2020A&A...639L..11I}
{Izzo}, L., {Auchettl}, K., {Hjorth}, J., {et~al.} 2020, \aap, 639, L11, \dodoi{10.1051/0004-6361/202038152}

\bibitem[{{Jin} {et~al.}(2013){Jin}, {Covino}, {Della Valle}, {Ferrero}, {Fugazza}, {Malesani}, {Melandri}, {Pian}, {Salvaterra}, {Bersier}, {Campana}, {Cano}, {Castro-Tirado}, {D'Avanzo}, {Fynbo}, {Gomboc}, {Gorosabel}, {Guidorzi}, {Haislip}, {Hjorth}, {Kobayashi}, {LaCluyze}, {Marconi}, {Mazzali}, {Mundell}, {Piranomonte}, {Reichart}, {S{\'a}nchez-Ram{\'\i}rez}, {Smith}, {Steele}, {Tagliaferri}, {Tanvir}, {Valenti}, {Vergani}, {Vestrand}, {Walker}, \& {Wo{\'z}niak}}]{2013ApJ...774..114J}
{Jin}, Z.-P., {Covino}, S., {Della Valle}, M., {et~al.} 2013, \apj, 774, 114, \dodoi{10.1088/0004-637X/774/2/114}

\bibitem[{{Kann} {et~al.}(2018){Kann}, {Schady}, {Olivares}, {Klose}, {Rossi}, {Perley}, {Zhang}, {Kr{\"u}hler}, {Greiner}, {Nicuesa Guelbenzu}, {Elliott}, {Knust}, {Cano}, {Filgas}, {Pian}, {Mazzali}, {Fynbo}, {Leloudas}, {Afonso}, {Delvaux}, {Graham}, {Rau}, {Schmidl}, {Schulze}, {Tanga}, {Updike}, \& {Varela}}]{2018AandA...617A.122K}
{Kann}, D.~A., {Schady}, P., {Olivares}, E.~F., {et~al.} 2018, \aap, 617, A122, \dodoi{10.1051/0004-6361/201731292}

\bibitem[{{Keshet} \& {Waxman}(2005)}]{2005PhRvL..94k1102K}
{Keshet}, U., \& {Waxman}, E. 2005, \prl, 94, 111102, \dodoi{10.1103/PhysRevLett.94.111102}

\bibitem[{{Klose} {et~al.}(2012){Klose}, {Greiner}, {Fynbo}, {Nicuesa Guelbenzu}, {Schmidl}, {Rau}, \& {Kruehler}}]{2012CBET.3200....1K}
{Klose}, S., {Greiner}, J., {Fynbo}, J., {et~al.} 2012, Central Bureau Electronic Telegrams, 3200, 1

\bibitem[{{Klose} {et~al.}(2019){Klose}, {Schmidl}, {Kann}, {Nicuesa Guelbenzu}, {Schulze}, {Greiner}, {Olivares E.}, {Kr{\"u}hler}, {Schady}, {Afonso}, {Filgas}, {Fynbo}, {Rau}, {Rossi}, {Takats}, {Tanga}, {Updike}, \& {Varela}}]{2019AandA...622A.138K}
{Klose}, S., {Schmidl}, S., {Kann}, D.~A., {et~al.} 2019, \aap, 622, A138, \dodoi{10.1051/0004-6361/201832728}

\bibitem[{{Kobayashi} \& {Sari}(2000)}]{2000ApJ...542..819K}
{Kobayashi}, S., \& {Sari}, R. 2000, \apj, 542, 819, \dodoi{10.1086/317021}

\bibitem[{{Kulkarni} {et~al.}(1998){Kulkarni}, {Frail}, {Wieringa}, {Ekers}, {Sadler}, {Wark}, {Higdon}, {Phinney}, \& {Bloom}}]{1998Natur.395..663K}
{Kulkarni}, S.~R., {Frail}, D.~A., {Wieringa}, M.~H., {et~al.} 1998, \nat, 395, 663, \dodoi{10.1038/27139}

\bibitem[{{Kusafuka} \& {Asano}(2025{\natexlab{a}})}]{2025MNRAS.536.1822K}
{Kusafuka}, Y., \& {Asano}, K. 2025{\natexlab{a}}, \mnras, 536, 1822, \dodoi{10.1093/mnras/stae2734}

\bibitem[{{Kusafuka} \& {Asano}(2025{\natexlab{b}})}]{2025mnras/staf879K}
---. 2025{\natexlab{b}}, \mnras, 540, 2098, \dodoi{10.1093/mnras/staf879}

\bibitem[{{Kusafuka} {et~al.}(2023){Kusafuka}, {Asano}, {Ohmura}, \& {Kawashima}}]{2023MNRAS.526..512K}
{Kusafuka}, Y., {Asano}, K., {Ohmura}, T., \& {Kawashima}, T. 2023, \mnras, 526, 512, \dodoi{10.1093/mnras/stad2804}

\bibitem[{{Kusafuka} {et~al.}(2025){Kusafuka}, {Obayashi}, {Asano}, \& {Yamazaki}}]{2025/mnras/staf1923K}
{Kusafuka}, Y., {Obayashi}, K., {Asano}, K., \& {Yamazaki}, R. 2025, \mnras, 544, 3115, \dodoi{10.1093/mnras/staf1923}

\bibitem[{{Lamb} \& {Kobayashi}(2017)}]{2017MNRAS.472.4953L}
{Lamb}, G.~P., \& {Kobayashi}, S. 2017, \mnras, 472, 4953, \dodoi{10.1093/mnras/stx2345}

\bibitem[{{Larsson} {et~al.}(2015){Larsson}, {Racusin}, \& {Burgess}}]{2015ApJ...800L..34L}
{Larsson}, J., {Racusin}, J.~L., \& {Burgess}, J.~M. 2015, \apjl, 800, L34, \dodoi{10.1088/2041-8205/800/2/L34}

\bibitem[{{Laskar} {et~al.}(2013){Laskar}, {Berger}, {Zauderer}, {Margutti}, {Soderberg}, {Chakraborti}, {Lunnan}, {Chornock}, {Chandra}, \& {Ray}}]{2013ApJ...776..119L}
{Laskar}, T., {Berger}, E., {Zauderer}, B.~A., {et~al.} 2013, \apj, 776, 119, \dodoi{10.1088/0004-637X/776/2/119}

\bibitem[{{LHAASO Collaboration} {et~al.}(2023){LHAASO Collaboration}, {Cao}, {Aharonian}, {An}, {Axikegu}, {Bai}, {Bai}, {Bao}, {Bastieri}, {Bi}, {Bi}, {Cai}, {Cao}, {Cao}, {Cao}, {Chang}, {Chang}, {Chen}, {Chen}, {Chen}, {Chen}, {Chen}, {Chen}, {Chen}, {Chen}, {Chen}, {Chen}, {Chen}, {Cheng}, {Cheng}, {Cheng}, {Cui}, {Cui}, {Cui}, {Dai}, {Dai}, {Danzengluobu}, {Della Volpe}, {Dong}, {Duan}, {Fan}, {Fan}, {Fang}, {Fang}, {Feng}, {Feng}, {Feng}, {Feng}, {Feng}, {Gao}, {Gao}, {Gao}, {Gao}, {Gao}, {Gao}, {Ge}, {Geng}, {Gong}, {Gou}, {Gu}, {Guo}, {Guo}, {Guo}, {Guo}, {Han}, {He}, {He}, {He}, {He}, {He}, {Heller}, {Hor}, {Hou}, {Hou}, {Hou}, {Hu}, {Hu}, {Hu}, {Huang}, {Huang}, {Huang}, {Huang}, {Huang}, {Ji}, {Jia}, {Jia}, {Jiang}, {Jiang}, {Jiang}, {Jin}, {Kang}, {Ke}, {Kuleshov}, {Kurinov}, {Li}, {Li}, {Li}, {Li}, {Li}, {Li}, {Li}, {Li}, {Li}, {Li}, {Li}, {Li}, {Li}, {Li}, {Li}, {Li}, {Li}, {Li}, {Li}, {Liang}, {Liang}, {Lin}, {Liu}, {Liu}, {Liu}, {Liu}, {Liu}, {Liu}, {Liu}, {Liu}, {Liu}, {Liu}, {Liu}, {Liu},
  {Liu}, {Liu}, {Liu}, {Liu}, {Long}, {Lu}, {Luo}, {Lv}, {Ma}, {Ma}, {Ma}, {Mao}, {Min}, {Mitthumsiri}, {Nan}, {Ou}, {Pang}, {Pattarakijwanich}, {Pei}, {Qi}, {Qi}, {Qiao}, {Qin}, {Ruffolo}, {Saiz}, {Shao}, {Shao}, {Shchegolev}, {Sheng}, {Song}, {Stenkin}, {Stepanov}, {Su}, {Sun}, {Sun}, {Sun}, {Tam}, {Tang}, {Tian}, {Wang}, {Wang}, {Wang}, {Wang}, {Wang}, {Wang}, {Wang}, {Wang}, {Wang}, {Wang}, {Wang}, {Wang}, {Wang}, {Wang}, {Wang}, {Wang}, {Wang}, {Wang}, {Wang}, {Wei}, {Wei}, {Wei}, {Wen}, {Wu}, {Wu}, {Wu}, {Wu}, {Wu}, {Xi}, {Xia}, {Xia}, {Xiang}, {Xiao}, {Xiao}, {Xin}, {Xin}, {Xing}, {Xiong}, {Xu}, {Xu}, {Xu}, {Xue}, {Yan}, {Yan}, {Yan}, {Yang}, {Yang}, {Yang}, {Yang}, {Yang}, {Yang}, {Yang}, {Yang}, {Yang}, {Yao}, {Ye}, {Yin}, {Yin}, {You}, {You}, {Yu}, {Yuan}, {Yue}, {Zeng}, {Zeng}, {Zeng}, {Zeng}, {Zhang}, {Zhang}, {Zhang}, {Zhang}, {Zhang}, {Zhang}, {Zhang}, {Zhang}, {Zhang}, {Zhang}, {Zhang}, {Zhang}, {Zhang}, {Zhang}, {Zhang}, {Zhang}, {Zhang}, {Zhang}, {Zhang}, {Zhao}, {Zhao}, {Zhao}, {Zhao},
  {Zhao}, {Zheng}, {Zhou}, {Zhou}, {Zhou}, {Zhou}, {Zhou}, {Zhou}, {Zhu}, {Zhu}, {Zhu}, {Zhu}, \& {Zuo}}]{2023Sci...380.1390L}
{LHAASO Collaboration}, {Cao}, Z., {Aharonian}, F., {et~al.} 2023, Science, 380, 1390, \dodoi{10.1126/science.adg9328}

\bibitem[{{Li} {et~al.}(2024){Li}, {Zhang}, {Huang}, \& {Xu}}]{2024ApJ...962..117L}
{Li}, X.-J., {Zhang}, Z.-B., {Huang}, Y.-F., \& {Xu}, F. 2024, \apj, 962, 117, \dodoi{10.3847/1538-4357/ad18a8}

\bibitem[{{Lipkin} {et~al.}(2004){Lipkin}, {Ofek}, {Gal-Yam}, {Leibowitz}, {Poznanski}, {Kaspi}, {Polishook}, {Kulkarni}, {Fox}, {Berger}, {Mirabal}, {Halpern}, {Bureau}, {Fathi}, {Price}, {Peterson}, {Frebel}, {Schmidt}, {Orosz}, {Fitzgerald}, {Bloom}, {van Dokkum}, {Bailyn}, {Buxton}, \& {Barsony}}]{2004ApJ...606..381L}
{Lipkin}, Y.~M., {Ofek}, E.~O., {Gal-Yam}, A., {et~al.} 2004, \apj, 606, 381, \dodoi{10.1086/383000}

\bibitem[{{MacFadyen} {et~al.}(2001){MacFadyen}, {Woosley}, \& {Heger}}]{2001ApJ...550..410M}
{MacFadyen}, A.~I., {Woosley}, S.~E., \& {Heger}, A. 2001, \apj, 550, 410, \dodoi{10.1086/319698}

\bibitem[{{Maeda}(2012)}]{2012ApJ...758...81M}
{Maeda}, K. 2012, \apj, 758, 81, \dodoi{10.1088/0004-637X/758/2/81}

\bibitem[{{Maeda}(2013)}]{2013ApJ...762L..24M}
---. 2013, \apjl, 762, L24, \dodoi{10.1088/2041-8205/762/2/L24}

\bibitem[{{Maeda} {et~al.}(2021){Maeda}, {Chandra}, {Matsuoka}, {Ryder}, {Moriya}, {Kuncarayakti}, {Lee}, {Kundu}, {Patnaude}, {Saito}, \& {Folatelli}}]{2021ApJ...918...34M}
{Maeda}, K., {Chandra}, P., {Matsuoka}, T., {et~al.} 2021, \apj, 918, 34, \dodoi{10.3847/1538-4357/ac0dbc}

\bibitem[{{Margutti} {et~al.}(2013){Margutti}, {Soderberg}, {Wieringa}, {Edwards}, {Chevalier}, {Morsony}, {Barniol Duran}, {Sironi}, {Zauderer}, {Milisavljevic}, {Kamble}, \& {Pian}}]{2013ApJ...778...18M}
{Margutti}, R., {Soderberg}, A.~M., {Wieringa}, M.~H., {et~al.} 2013, \apj, 778, 18, \dodoi{10.1088/0004-637X/778/1/18}

\bibitem[{{Marongiu} {et~al.}(2019){Marongiu}, {Guidorzi}, {Margutti}, {Coppejans}, {Martone}, \& {Kamble}}]{2019ApJ...879...89M}
{Marongiu}, M., {Guidorzi}, C., {Margutti}, R., {et~al.} 2019, \apj, 879, 89, \dodoi{10.3847/1538-4357/ab25ef}

\bibitem[{{Maselli} {et~al.}(2014){Maselli}, {Melandri}, {Nava}, {Mundell}, {Kawai}, {Campana}, {Covino}, {Cummings}, {Cusumano}, {Evans}, {Ghirlanda}, {Ghisellini}, {Guidorzi}, {Kobayashi}, {Kuin}, {La Parola}, {Mangano}, {Oates}, {Sakamoto}, {Serino}, {Virgili}, {Zhang}, {Barthelmy}, {Beardmore}, {Bernardini}, {Bersier}, {Burrows}, {Calderone}, {Capalbi}, {Chiang}, {D'Avanzo}, {D'Elia}, {De Pasquale}, {Fugazza}, {Gehrels}, {Gomboc}, {Harrison}, {Hanayama}, {Japelj}, {Kennea}, {Kopac}, {Kouveliotou}, {Kuroda}, {Levan}, {Malesani}, {Marshall}, {Nousek}, {O'Brien}, {Osborne}, {Pagani}, {Page}, {Page}, {Perri}, {Pritchard}, {Romano}, {Saito}, {Sbarufatti}, {Salvaterra}, {Steele}, {Tanvir}, {Vianello}, {Weigand}, {Wiersema}, {Yatsu}, {Yoshii}, \& {Tagliaferri}}]{2014Sci...343...48M}
{Maselli}, A., {Melandri}, A., {Nava}, L., {et~al.} 2014, Science, 343, 48, \dodoi{10.1126/science.1242279}

\bibitem[{{Matsuoka} {et~al.}(2024){Matsuoka}, {Kimura}, {Maeda}, \& {Tanaka}}]{2024ApJ...960...70M}
{Matsuoka}, T., {Kimura}, S.~S., {Maeda}, K., \& {Tanaka}, M. 2024, \apj, 960, 70, \dodoi{10.3847/1538-4357/ad096c}

\bibitem[{{Matsuoka} {et~al.}(2025){Matsuoka}, {Maeda}, {Kimura}, \& {Tanaka}}]{2025arXiv250506609M}
{Matsuoka}, T., {Maeda}, K., {Kimura}, S.~S., \& {Tanaka}, M. 2025, arXiv e-prints, arXiv:2505.06609, \dodoi{10.48550/arXiv.2505.06609}

\bibitem[{{Matsuoka} {et~al.}(2019){Matsuoka}, {Maeda}, {Lee}, \& {Yasuda}}]{2019ApJ...885...41M}
{Matsuoka}, T., {Maeda}, K., {Lee}, S.-H., \& {Yasuda}, H. 2019, \apj, 885, 41, \dodoi{10.3847/1538-4357/ab4421}

\bibitem[{{Melandri} {et~al.}(2012){Melandri}, {Pian}, {Ferrero}, {D'Elia}, {Walker}, {Ghirlanda}, {Covino}, {Amati}, {D'Avanzo}, {Mazzali}, {Della Valle}, {Guidorzi}, {Antonelli}, {Bernardini}, {Bersier}, {Bufano}, {Campana}, {Castro-Tirado}, {Chincarini}, {Deng}, {Filippenko}, {Fugazza}, {Ghisellini}, {Kouveliotou}, {Maeda}, {Marconi}, {Masetti}, {Nomoto}, {Palazzi}, {Patat}, {Piranomonte}, {Salvaterra}, {Saviane}, {Starling}, {Tagliaferri}, {Tanaka}, \& {Vergani}}]{2012AandA...547A..82M}
{Melandri}, A., {Pian}, E., {Ferrero}, P., {et~al.} 2012, \aap, 547, A82, \dodoi{10.1051/0004-6361/201219879}

\bibitem[{{Melandri} {et~al.}(2019){Melandri}, {Malesani}, {Izzo}, {Japelj}, {Vergani}, {Schady}, {Sagu{\'e}s Carracedo}, {de Ugarte Postigo}, {Anderson}, {Barbarino}, {Bolmer}, {Breeveld}, {Calissendorff}, {Campana}, {Cano}, {Carini}, {Covino}, {D'Avanzo}, {D'Elia}, {della Valle}, {De Pasquale}, {Fynbo}, {Gromadzki}, {Hammer}, {Hartmann}, {Heintz}, {Inserra}, {Jakobsson}, {Kann}, {Kotilainen}, {Maguire}, {Masetti}, {Nicholl}, {Olivares E}, {Pugliese}, {Rossi}, {Salvaterra}, {Sollerman}, {Stone}, {Tagliaferri}, {Tomasella}, {Th{\"o}ne}, {Xu}, \& {Young}}]{2019MNRAS.490.5366M}
{Melandri}, A., {Malesani}, D.~B., {Izzo}, L., {et~al.} 2019, \mnras, 490, 5366, \dodoi{10.1093/mnras/stz2900}

\bibitem[{{Melandri} {et~al.}(2022){Melandri}, {Izzo}, {Pian}, {Malesani}, {Della Valle}, {Rossi}, {D'Avanzo}, {Guetta}, {Mazzali}, {Benetti}, {Masetti}, {Palazzi}, {Savaglio}, {Amati}, {Antonelli}, {Ashall}, {Bernardini}, {Campana}, {Carini}, {Covino}, {D'Elia}, {de Ugarte Postigo}, {De Pasquale}, {Filippenko}, {Fruchter}, {Fynbo}, {Giunta}, {Hartmann}, {Jakobsson}, {Japelj}, {Jonker}, {Kann}, {Lamb}, {Levan}, {Martin-Carrillo}, {M{\o}ller}, {Piranomonte}, {Pugliese}, {Salvaterra}, {Schulze}, {Starling}, {Stella}, {Tagliaferri}, {Tanvir}, \& {Watson}}]{2022AandA...659A..39M}
{Melandri}, A., {Izzo}, L., {Pian}, E., {et~al.} 2022, \aap, 659, A39, \dodoi{10.1051/0004-6361/202141788}

\bibitem[{{Mimica} {et~al.}(2009){Mimica}, {Giannios}, \& {Aloy}}]{2009AandA...494..879M}
{Mimica}, P., {Giannios}, D., \& {Aloy}, M.~A. 2009, \aap, 494, 879, \dodoi{10.1051/0004-6361:200810756}

\bibitem[{{Nugis} \& {Lamers}(2000)}]{2000AandA...360..227N}
{Nugis}, T., \& {Lamers}, H.~J.~G.~L.~M. 2000, \aap, 360, 227

\bibitem[{{Obayashi} {et~al.}(2024){Obayashi}, {Toriyama}, {Murakoshi}, {Sato}, {Tanaka}, {Sakamoto}, \& {Yamazaki}}]{2024JHEAp..41....1O}
{Obayashi}, K., {Toriyama}, A., {Murakoshi}, M., {et~al.} 2024, Journal of High Energy Astrophysics, 41, 1, \dodoi{10.1016/j.jheap.2023.12.001}

\bibitem[{{Omand} {et~al.}(2025){Omand}, {Sarin}, \& {Lamb}}]{2025MNRAS.539.1908O}
{Omand}, C. M.~B., {Sarin}, N., \& {Lamb}, G.~P. 2025, \mnras, 539, 1908, \dodoi{10.1093/mnras/staf565}

\bibitem[{{Ostriker} \& {McKee}(1988)}]{1988RvMP...60....1O}
{Ostriker}, J.~P., \& {McKee}, C.~F. 1988, Reviews of Modern Physics, 60, 1, \dodoi{10.1103/RevModPhys.60.1}

\bibitem[{{Palliyaguru} {et~al.}(2021){Palliyaguru}, {Corsi}, {P{\'e}rez-Torres}, {Varenius}, \& {Van Eerten}}]{2021ApJ...910...16P}
{Palliyaguru}, N.~T., {Corsi}, A., {P{\'e}rez-Torres}, M., {Varenius}, E., \& {Van Eerten}, H. 2021, \apj, 910, 16, \dodoi{10.3847/1538-4357/abe1c9}

\bibitem[{{Pang} \& {Dai}(2024)}]{2024MNRAS.528.2066P}
{Pang}, S.-L., \& {Dai}, Z.-G. 2024, \mnras, 528, 2066, \dodoi{10.1093/mnras/stae197}

\bibitem[{{Perley} {et~al.}(2014){Perley}, {Cenko}, {Corsi}, {Tanvir}, {Levan}, {Kann}, {Sonbas}, {Wiersema}, {Zheng}, {Zhao}, {Bai}, {Bremer}, {Castro-Tirado}, {Chang}, {Clubb}, {Frail}, {Fruchter}, {G{\"o}{\u{g}}{\"u}{\c{s}}}, {Greiner}, {G{\"u}ver}, {Horesh}, {Filippenko}, {Klose}, {Mao}, {Morgan}, {Pozanenko}, {Schmidl}, {Stecklum}, {Tanga}, {Volnova}, {Volvach}, {Wang}, {Winters}, \& {Xin}}]{2014ApJ...781...37P}
{Perley}, D.~A., {Cenko}, S.~B., {Corsi}, A., {et~al.} 2014, \apj, 781, 37, \dodoi{10.1088/0004-637X/781/1/37}

\bibitem[{{Price} {et~al.}(2002){Price}, {Berger}, {Reichart}, {Kulkarni}, {Yost}, {Subrahmanyan}, {Wark}, {Wieringa}, {Frail}, {Bailey}, {Boyle}, {Corbett}, {Gunn}, {Ryder}, {Seymour}, {Koviak}, {McCarthy}, {Phillips}, {Axelrod}, {Bloom}, {Djorgovski}, {Fox}, {Galama}, {Harrison}, {Hurley}, {Sari}, {Schmidt}, {Brown}, {Cline}, {Frontera}, {Guidorzi}, \& {Montanari}}]{2002ApJ...572L..51P}
{Price}, P.~A., {Berger}, E., {Reichart}, D.~E., {et~al.} 2002, \apjl, 572, L51, \dodoi{10.1086/341552}

\bibitem[{{Ramirez-Ruiz} {et~al.}(2002){Ramirez-Ruiz}, {Celotti}, \& {Rees}}]{2002MNRAS.337.1349R}
{Ramirez-Ruiz}, E., {Celotti}, A., \& {Rees}, M.~J. 2002, \mnras, 337, 1349, \dodoi{10.1046/j.1365-8711.2002.05995.x}

\bibitem[{{Ramirez-Ruiz} {et~al.}(2005){Ramirez-Ruiz}, {Granot}, {Kouveliotou}, {Woosley}, {Patel}, \& {Mazzali}}]{2005ApJ...625L..91R}
{Ramirez-Ruiz}, E., {Granot}, J., {Kouveliotou}, C., {et~al.} 2005, \apjl, 625, L91, \dodoi{10.1086/431237}

\bibitem[{{Resmi} {et~al.}(2012){Resmi}, {Misra}, {J{\'o}hannesson}, {Castro Tirado}, {Gorosabel}, {Jel{\'\i}nek}, {Bhattacharya}, {Kub{\'a}nek}, {Anupama}, {Sota}, {Sahu}, {de Ugarte Postigo}, {Pandey}, {S{\'a}nchez Ram{\'\i}rez}, {Bremer}, \& {Sagar}}]{2012MNRAS.427..288R}
{Resmi}, L., {Misra}, K., {J{\'o}hannesson}, G., {et~al.} 2012, \mnras, 427, 288, \dodoi{10.1111/j.1365-2966.2012.21713.x}

\bibitem[{{Ressler} \& {Laskar}(2017)}]{2017ApJ...845..150R}
{Ressler}, S.~M., \& {Laskar}, T. 2017, \apj, 845, 150, \dodoi{10.3847/1538-4357/aa8268}

\bibitem[{{Rhoads}(1999)}]{1999ApJ...525..737R}
{Rhoads}, J.~E. 1999, \apj, 525, 737, \dodoi{10.1086/307907}

\bibitem[{{Rose} {et~al.}(2024){Rose}, {Horesh}, {Murphy}, {Kaplan}, {Sfaradi}, {Ryder}, {Aloisi}, {Dobie}, {Driessen}, {Fender}, {Green}, {Leung}, {Lenc}, {Qiu}, \& {Williams-Baldwin}}]{2024MNRAS.534.3853R}
{Rose}, K., {Horesh}, A., {Murphy}, T., {et~al.} 2024, \mnras, 534, 3853, \dodoi{10.1093/mnras/stae2289}

\bibitem[{{Rossi} {et~al.}(2002){Rossi}, {Lazzati}, \& {Rees}}]{2002MNRAS.332..945R}
{Rossi}, E., {Lazzati}, D., \& {Rees}, M.~J. 2002, \mnras, 332, 945, \dodoi{10.1046/j.1365-8711.2002.05363.x}

\bibitem[{{Ryan} {et~al.}(2020){Ryan}, {van Eerten}, {Piro}, \& {Troja}}]{2020ApJ...896..166R}
{Ryan}, G., {van Eerten}, H., {Piro}, L., \& {Troja}, E. 2020, \apj, 896, 166, \dodoi{10.3847/1538-4357/ab93cf}

\bibitem[{{Rybicki} \& {Lightman}(1986)}]{1986rpa..book.....R}
{Rybicki}, G.~B., \& {Lightman}, A.~P. 1986, {Radiative Processes in Astrophysics} (New York, NY: Wiley), \dodoi{10.1002/9783527618170}

\bibitem[{{Salas} {et~al.}(2013){Salas}, {Bauer}, {Stockdale}, \& {Prieto}}]{2013MNRAS.428.1207S}
{Salas}, P., {Bauer}, F.~E., {Stockdale}, C., \& {Prieto}, J.~L. 2013, \mnras, 428, 1207, \dodoi{10.1093/mnras/sts104}

\bibitem[{{Sari} \& {Piran}(1995)}]{1995ApJ...455L.143S}
{Sari}, R., \& {Piran}, T. 1995, \apjl, 455, L143, \dodoi{10.1086/309835}

\bibitem[{{Sari} {et~al.}(1998){Sari}, {Piran}, \& {Narayan}}]{1998ApJ...497L..17S}
{Sari}, R., {Piran}, T., \& {Narayan}, R. 1998, \apjl, 497, L17, \dodoi{10.1086/311269}

\bibitem[{{Sawada} \& {Ashida}(2025)}]{2025ApJ...982...93S}
{Sawada}, R., \& {Ashida}, Y. 2025, \apj, 982, 93, \dodoi{10.3847/1538-4357/adb721}

\bibitem[{{Schroeder} {et~al.}(2024){Schroeder}, {Rhodes}, {Laskar}, {Nugent}, {Rouco Escorial}, {Rastinejad}, {Fong}, {van der Horst}, {Veres}, {Alexander}, {Andersson}, {Berger}, {Blanchard}, {Chastain}, {Christensen}, {Fender}, {Green}, {Groot}, {Heywood}, {Horesh}, {Izzo}, {Kilpatrick}, {K{\"o}rding}, {Lien}, {Malesani}, {McBride}, {Mooley}, {Rowlinson}, {Sears}, {Stappers}, {Tanvir}, {Vergani}, {Wijers}, {Williams-Baldwin}, \& {Woudt}}]{2024ApJ...970..139S}
{Schroeder}, G., {Rhodes}, L., {Laskar}, T., {et~al.} 2024, \apj, 970, 139, \dodoi{10.3847/1538-4357/ad49ab}

\bibitem[{{Shivvers} {et~al.}(2017){Shivvers}, {Modjaz}, {Zheng}, {Liu}, {Filippenko}, {Silverman}, {Matheson}, {Pastorello}, {Graur}, {Foley}, {Chornock}, {Smith}, {Leaman}, \& {Benetti}}]{2017PASP..129e4201S}
{Shivvers}, I., {Modjaz}, M., {Zheng}, W., {et~al.} 2017, \pasp, 129, 054201, \dodoi{10.1088/1538-3873/aa54a6}

\bibitem[{{Singer} {et~al.}(2013){Singer}, {Cenko}, {Kasliwal}, {Perley}, {Ofek}, {Brown}, {Nugent}, {Kulkarni}, {Corsi}, {Frail}, {Bellm}, {Mulchaey}, {Arcavi}, {Barlow}, {Bloom}, {Cao}, {Gehrels}, {Horesh}, {Masci}, {McEnery}, {Rau}, {Surace}, \& {Yaron}}]{2013ApJ...776L..34S}
{Singer}, L.~P., {Cenko}, S.~B., {Kasliwal}, M.~M., {et~al.} 2013, \apjl, 776, L34, \dodoi{10.1088/2041-8205/776/2/L34}

\bibitem[{{Sironi} {et~al.}(2015){Sironi}, {Petropoulou}, \& {Giannios}}]{2015MNRAS.450..183S}
{Sironi}, L., {Petropoulou}, M., \& {Giannios}, D. 2015, \mnras, 450, 183, \dodoi{10.1093/mnras/stv641}

\bibitem[{{Sironi} \& {Spitkovsky}(2011)}]{2011ApJ...726...75S}
{Sironi}, L., \& {Spitkovsky}, A. 2011, \apj, 726, 75, \dodoi{10.1088/0004-637X/726/2/75}

\bibitem[{{Soderberg} {et~al.}(2006){Soderberg}, {Nakar}, {Berger}, \& {Kulkarni}}]{2006ApJ...638..930S}
{Soderberg}, A.~M., {Nakar}, E., {Berger}, E., \& {Kulkarni}, S.~R. 2006, \apj, 638, 930, \dodoi{10.1086/499121}

\bibitem[{{Soderberg} {et~al.}(2010){Soderberg}, {Chakraborti}, {Pignata}, {Chevalier}, {Chandra}, {Ray}, {Wieringa}, {Copete}, {Chaplin}, {Connaughton}, {Barthelmy}, {Bietenholz}, {Chugai}, {Stritzinger}, {Hamuy}, {Fransson}, {Fox}, {Levesque}, {Grindlay}, {Challis}, {Foley}, {Kirshner}, {Milne}, \& {Torres}}]{2010Natur.463..513S}
{Soderberg}, A.~M., {Chakraborti}, S., {Pignata}, G., {et~al.} 2010, \nat, 463, 513, \dodoi{10.1038/nature08714}

\bibitem[{{Sparre} {et~al.}(2011){Sparre}, {Sollerman}, {Fynbo}, {Malesani}, {Goldoni}, {de Ugarte Postigo}, {Covino}, {D'Elia}, {Flores}, {Hammer}, {Hjorth}, {Jakobsson}, {Kaper}, {Leloudas}, {Levan}, {Milvang-Jensen}, {Schulze}, {Tagliaferri}, {Tanvir}, {Watson}, {Wiersema}, \& {Wijers}}]{2011ApJ...735L..24S}
{Sparre}, M., {Sollerman}, J., {Fynbo}, J.~P.~U., {et~al.} 2011, \apjl, 735, L24, \dodoi{10.1088/2041-8205/735/1/L24}

\bibitem[{{Srinivasaragavan} {et~al.}(2023){Srinivasaragavan}, {O'Connor}, {Cenko}, {Dittmann}, {Yang}, {Sollerman}, {Anupama}, {Barway}, {Bhalerao}, {Kumar}, {Swain}, {Hammerstein}, {Holt}, {Anand}, {Andreoni}, {Coughlin}, {Dichiara}, {Gal-Yam}, {Miller}, {Soon}, {Soria}, {Durbak}, {Gillanders}, {Laha}, {Moore}, {Ragosta}, \& {Troja}}]{2023ApJ...949L..39S}
{Srinivasaragavan}, G.~P., {O'Connor}, B., {Cenko}, S.~B., {et~al.} 2023, \apjl, 949, L39, \dodoi{10.3847/2041-8213/accf97}

\bibitem[{{Suzuki} \& {Maeda}(2018)}]{2018MNRAS.478..110S}
{Suzuki}, A., \& {Maeda}, K. 2018, \mnras, 478, 110, \dodoi{10.1093/mnras/sty999}

\bibitem[{{Thomas} {et~al.}(2017){Thomas}, {Moharana}, \& {Razzaque}}]{2017PhRvD..96j3004T}
{Thomas}, J.~K., {Moharana}, R., \& {Razzaque}, S. 2017, \prd, 96, 103004, \dodoi{10.1103/PhysRevD.96.103004}

\bibitem[{{Thomsen} {et~al.}(2004){Thomsen}, {Hjorth}, {Watson}, {Gorosabel}, {Fynbo}, {Jensen}, {Andersen}, {Dall}, {Rasmussen}, {Bruntt}, {Laurikainen}, {Augusteijn}, {Pursimo}, {Germany}, {Jakobsson}, \& {Pedersen}}]{2004AandA...419L..21T}
{Thomsen}, B., {Hjorth}, J., {Watson}, D., {et~al.} 2004, \aap, 419, L21, \dodoi{10.1051/0004-6361:20040133}

\bibitem[{{Troja} {et~al.}(2018){Troja}, {Piro}, {Ryan}, {van Eerten}, {Ricci}, {Wieringa}, {Lotti}, {Sakamoto}, \& {Cenko}}]{2018MNRAS.478L..18T}
{Troja}, E., {Piro}, L., {Ryan}, G., {et~al.} 2018, \mnras, 478, L18, \dodoi{10.1093/mnrasl/sly061}

\bibitem[{{Troja} {et~al.}(2019){Troja}, {van Eerten}, {Ryan}, {Ricci}, {Burgess}, {Wieringa}, {Piro}, {Cenko}, \& {Sakamoto}}]{2019MNRAS.489.1919T}
{Troja}, E., {van Eerten}, H., {Ryan}, G., {et~al.} 2019, \mnras, 489, 1919, \dodoi{10.1093/mnras/stz2248}

\bibitem[{Truelove \& McKee(1999)}]{1999ApJS..120..299T}
Truelove, J.~K., \& McKee, C.~F. 1999, The Astrophysical Journal Supplement Series, 120, 299, \dodoi{10.1086/313176}

\bibitem[{{Urata} {et~al.}(2015){Urata}, {Huang}, {Yamazaki}, \& {Sakamoto}}]{2015ApJ...806..222U}
{Urata}, Y., {Huang}, K., {Yamazaki}, R., \& {Sakamoto}, T. 2015, \apj, 806, 222, \dodoi{10.1088/0004-637X/806/2/222}

\bibitem[{{van der Horst} {et~al.}(2011){van der Horst}, {Kamble}, {Paragi}, {Sage}, {Pal}, {Taylor}, {Kouveliotou}, {Granot}, {Ramirez-Ruiz}, {Ishwara-Chandra}, {Oosterloo}, {Wijers}, {Wiersema}, {Strom}, {Bhattacharya}, {Rol}, {Starling}, {Curran}, \& {Garrett}}]{2011ApJ...726...99V}
{van der Horst}, A.~J., {Kamble}, A.~P., {Paragi}, Z., {et~al.} 2011, \apj, 726, 99, \dodoi{10.1088/0004-637X/726/2/99}

\bibitem[{{van Eerten}(2014)}]{2014MNRAS.442.3495V}
{van Eerten}, H. 2014, \mnras, 442, 3495, \dodoi{10.1093/mnras/stu1025}

\bibitem[{{van Eerten} \& {MacFadyen}(2012)}]{2012ApJ...751..155V}
{van Eerten}, H.~J., \& {MacFadyen}, A.~I. 2012, \apj, 751, 155, \dodoi{10.1088/0004-637X/751/2/155}

\bibitem[{{van Marle} {et~al.}(2006){van Marle}, {Langer}, {Achterberg}, \& {Garc{\'\i}a-Segura}}]{2006AandA...460..105V}
{van Marle}, A.~J., {Langer}, N., {Achterberg}, A., \& {Garc{\'\i}a-Segura}, G. 2006, \aap, 460, 105, \dodoi{10.1051/0004-6361:20065709}

\bibitem[{{Vergani} {et~al.}(2011){Vergani}, {Flores}, {Covino}, {Fugazza}, {Gorosabel}, {Levan}, {Puech}, {Salvaterra}, {Tello}, {de Ugarte Postigo}, {D'Avanzo}, {D'Elia}, {Fern{\'a}ndez}, {Fynbo}, {Ghirlanda}, {Jel{\'\i}nek}, {Lundgren}, {Malesani}, {Palazzi}, {Piranomonte}, {Rodrigues}, {S{\'a}nchez-Ram{\'\i}rez}, {Terr{\'o}n}, {Th{\"o}ne}, {Antonelli}, {Campana}, {Castro-Tirado}, {Goldoni}, {Hammer}, {Hjorth}, {Jakobsson}, {Kaper}, {Melandri}, {Milvang-Jensen}, {Sollerman}, {Tagliaferri}, {Tanvir}, {Wiersema}, \& {Wijers}}]{2011AandA...535A.127V}
{Vergani}, S.~D., {Flores}, H., {Covino}, S., {et~al.} 2011, \aap, 535, A127, \dodoi{10.1051/0004-6361/201117726}

\bibitem[{{Wang} {et~al.}(2024){Wang}, {Dastidar}, {Giannios}, \& {Duffell}}]{2024ApJS..273...17W}
{Wang}, H., {Dastidar}, R.~G., {Giannios}, D., \& {Duffell}, P.~C. 2024, \apjs, 273, 17, \dodoi{10.3847/1538-4365/ad4d9d}

\bibitem[{{Wang} {et~al.}(2025){Wang}, {Zhou}, {Fan}, \& {Wei}}]{2025ApJ...990..110W}
{Wang}, H., {Zhou}, H., {Fan}, Y.-Z., \& {Wei}, D.-M. 2025, \apj, 990, 110, \dodoi{10.3847/1538-4357/adf1a2}

\bibitem[{{Wang} \& {Wheeler}(1998)}]{1998ApJ...504L..87W}
{Wang}, L., \& {Wheeler}, J.~C. 1998, \apjl, 504, L87, \dodoi{10.1086/311580}

\bibitem[{{Wang} {et~al.}(2019){Wang}, {Rueda}, {Ruffini}, {Becerra}, {Bianco}, {Becerra}, {Li}, \& {Karlica}}]{2019ApJ...874...39W}
{Wang}, Y., {Rueda}, J.~A., {Ruffini}, R., {et~al.} 2019, \apj, 874, 39, \dodoi{10.3847/1538-4357/ab04f8}

\bibitem[{{Warren} {et~al.}(2018){Warren}, {Barkov}, {Ito}, {Nagataki}, \& {Laskar}}]{2018MNRAS.480.4060W}
{Warren}, D.~C., {Barkov}, M.~V., {Ito}, H., {Nagataki}, S., \& {Laskar}, T. 2018, \mnras, 480, 4060, \dodoi{10.1093/mnras/sty2138}

\bibitem[{{Waxman}(1997)}]{1997ApJ...491L..19W}
{Waxman}, E. 1997, \apjl, 491, L19, \dodoi{10.1086/311057}

\bibitem[{{Waxman}(2004)}]{2004ApJ...602..886W}
---. 2004, \apj, 602, 886, \dodoi{10.1086/381230}

\bibitem[{{Wei} {et~al.}(2023){Wei}, {Zhang}, \& {Murase}}]{2023MNRAS.524.6004W}
{Wei}, Y., {Zhang}, B.~T., \& {Murase}, K. 2023, \mnras, 524, 6004, \dodoi{10.1093/mnras/stad2122}

\bibitem[{{Woods} \& {Loeb}(1999)}]{1999ApJ...523..187W}
{Woods}, E., \& {Loeb}, A. 1999, \apj, 523, 187, \dodoi{10.1086/307738}

\bibitem[{{Woosley} \& {Bloom}(2006)}]{2006ARAandA..44..507W}
{Woosley}, S.~E., \& {Bloom}, J.~S. 2006, \araa, 44, 507, \dodoi{10.1146/annurev.astro.43.072103.150558}

\bibitem[{{Young} {et~al.}(2010){Young}, {Smartt}, {Valenti}, {Pastorello}, {Benetti}, {Benn}, {Bersier}, {Botticella}, {Corradi}, {Harutyunyan}, {Hrudkova}, {Hunter}, {Mattila}, {de Mooij}, {Navasardyan}, {Snellen}, {Tanvir}, \& {Zampieri}}]{2010AandA...512A..70Y}
{Young}, D.~R., {Smartt}, S.~J., {Valenti}, S., {et~al.} 2010, \aap, 512, A70, \dodoi{10.1051/0004-6361/200913004}

\bibitem[{{Zhang}(2018)}]{2018pgrb.book.....Z}
{Zhang}, B. 2018, {The Physics of Gamma-Ray Bursts}, \dodoi{10.1017/9781139226530}

\bibitem[{{Zhang} {et~al.}(2012){Zhang}, {Fan}, {Shen}, {Xu}, {Zhang}, {Wei}, {Burrows}, {Zhang}, \& {Gehrels}}]{2012ApJ...756..190Z}
{Zhang}, B.-B., {Fan}, Y.-Z., {Shen}, R.-F., {et~al.} 2012, \apj, 756, 190, \dodoi{10.1088/0004-637X/756/2/190}

\bibitem[{{Zhang} {et~al.}(2016){Zhang}, {Huang}, \& {Zong}}]{2016ApJ...823..156Z}
{Zhang}, Q., {Huang}, Y.~F., \& {Zong}, H.~S. 2016, \apj, 823, 156, \dodoi{10.3847/0004-637X/823/2/156}

\end{thebibliography}
\bibliographystyle{aasjournal}

\end{document}